\documentclass[aps,onecolumn,showpacs,showkeys,nofootinbib]{revtex4}
\usepackage{epsfig}
\usepackage{amsmath}
\usepackage{amsfonts}
\usepackage{amssymb}
\usepackage{graphicx}
\usepackage{colordvi}
\usepackage{subfigure}
\begin{document}
\begin{flushleft}
\end{flushleft}

\title{Precession shift in curvature based Extended Theories of Gravity and Quintessence fields}

\author{A. Capolupo$^{1,2}$\footnote{ e-mail address: capolupo@sa.infn.it}, G. Lambiase$^{1,2}$\footnote{e-mail address: lambiase@sa.infn.it}, A. Tedesco $^{1,2}$\footnote{e-mail address: atedesco@unisa.it}}


\affiliation{$^1$Dipartimento di Fisica ``E.R. Caianiello'',  Universit\`{a} degli Studi di Salerno, via G. Paolo II, I  - 84084 Fisciano, Italy}
\affiliation{$^2$Istituto Nazionale di  Fisica Nucleare (INFN), Gruppo collegato di Salerno, via G. Paolo II, I - 84084 Fisciano, Italy}

\begin{abstract}
In this paper we constrain the sizes of hypothetical new weak forces by making use the data coming from the precession of Planets. We consider the weak field approximation of Scalar-Tensor Fourth Order Gravity (STFOG), which include several models of modified gravity. The form of the corrections to the Newtonian potential if of the form of Yukawa-like potential (5th force), i.e. $V(r)=\alpha \frac{e^{-\beta r}}{r}$, where $\alpha$ is the parameter related to the strength of the potential, and $\beta$ to the range of the force. The present data on periastron advance allow to infer a constraint on the free parameter of the gravitational models. Moreover, the Non-Commutative Spectral Gravity (NCSG) is also studied, being a particular case STFOG. Here we show that the precession shift of Planet allows to improve the bounds on parameter $\beta$ by several orders of magnitude. Finally such an analysis is studied to the case of power-like potential, referring in particular to deformation of the Schwarzschild geometry induced by a quintessence field, responsible of the present accelerated phase of the Universe.
\end{abstract}
\date{\today}
\maketitle

\section{Introduction}

According to recent observations, the picture of the present Universe is that it is spatially flat and undergoing a period of accelerated expansion
\cite{riess,ast,clo,spe,carrol,sahini}. To dynamically address such a picture new ingredients must be necessarily introduced, the dark energy at and the dark matter. The former acts at cosmological scales and is related to the accelerated expansion of the Universe, the latter acts at galactic and extragalactic scales and is responsible for the clustering of structure. Many efforts and attempts have been done to explain the origin and nature of these two ingredients, without reaching an universal consensus. A possibility is related to the Extended Theories of Gravity (ETG) \cite{Felix} that allow to explain, on a pure gravitational setting and without the needed to invoke any exotic matter, the galactic rotation curves and the cosmic acceleration (see for example \cite{Nojiri,vasilis, felice}).
In the framework of ETG, hence the gravitational interaction acts differently at different scales, while the results of GR at Solar System scales are preserved. Therefore, GR is a particular case of a more extended class of theories. From a conceptual viewpoint, there is no \emph{a priori} reason to restrict the gravitational action to a linear function of the Ricci scalar minimally coupled to matter, as for the Hilbert-Einstein action \cite{mag-fer-fra}.

It must be mentioned that modifications of GR are also motivated by the fact that Einstein's theory of gravity breaks down in the UV \cite{Capozziello:2011et, Will:2018bme}. Such deviations from General Relativity occurs in several frameworks, such as Brans-Dicke and scalar-tensor (ST) theories~\cite{Perivolaropoulos:2009ak,Hohmann:2013rba,Jarv:2014hma}, braneworld theories~\cite{Nojiri:2002wn,Bronnikov:2006jy,Kaminski:2009dh,Benichou:2011dx, Guo:2014bxa,Donini:2016kgu}, higher order invariants such as $f(R)$ and $f(\phi,\,R,\,R^2, R_{\mu\nu}R^{\mu\nu}, \Box R)$ (which correspond to Einstein's gravity plus one or multiple conformally coupled scalar fields~\cite{Teyssandier:1983zz, Maeda:1988ab, Wands:1993uu, Schmidt:2001ac})  \cite{Berry:2011pb,Capozziello:2014mea, Lambiase:2015yia,Schellstede:2016ldu,NatureSci}, noncommutative geometry~\cite{Lambiase:2013dai}, and compactified extra dimension/Kaluza-Klein models~\cite{ArkaniHamed:1998rs,ArkaniHamed:1998nn,Antoniadis:1998ig,Floratos:1999bv,Kehagias:1999my,Perivolaropoulos:2002pn}.
Moreover, they can also be generated from higher-order terms in the curvature invariants, nonminimal couplings to the background geometry in the Hilbert-Einstein Lagrangian~\cite{Birrell:1982ix, Gasperini:1991ak, Vilkovisky:1992pb, Nojiri:2006ri}.  Additional terms into the action of gravity may also come from string loop effects~\cite{Damour:1994ya}, dilaton fields in string cosmology~\cite{Gasperini:1994xg}, and nonlocally  modified  gravity induced by  quantum  loop  corrections~\cite{Deser:2007jk}. Moreover, ETG are not only curvature based but  can involve also other formulations like  affine connections independent of the metric, as in the case of  metric-affine gravity  \cite{ext1} or   purely affine gravity \cite{ext2}. Metric-affine theories are also the Poincar\'e gauge gravity \cite{ext3}, the teleparallel gravity based on the Weitzenb\"ock connection \cite{ext4,ext5}, the symmetric teleparallel gravity \cite{ext6}. The debate on the identification of variables describing the gravitational field is still open and it is a very active  research area. In this paper, we are going to consider curvature based extended theories, confining ourselves to the case of metric theories. We analyze the weak field limit of Scalar-Tensor Fourth Order Gravity models and the effects of these corrections to the periastron advance of Planets. We consider several theories where higher-order curvature invariants and a scalar field are included. In particular, we study the Non-Commutative Spectral Gravity and  the effect of the Quintessence field present around a Schwarzschild Black Hole. Then, we derive a lower bound on the adiabatic index of equation of state.

The Newtonian limit of some models of ETG have been studied in \cite{PRD1,mio2}, while the Minkowskian limit in \cite{quadrupolo,Sta1,FOG_CGL2,FOGGW,leo,capriolo1,capriolo2}. Natural candidates for experimentally testing ETG are the galactic rotation curves, stellar systems and gravitational lensing \cite{ BHL, BHL1, stabile_scelza,stabile_scelza2,stabstab,stabile_stabile_cap} (see also \cite{lv,Lambiase:2016bjy4,LambMohantySta}). In this perspective, corrections to GR were already considered by several authors
\cite{weyl_2,edd,lan,pauli,bach,buc,bic,FOG_CGL,FOG_CGL2,FOG_CGL3,FOG_CGL4,FOG_CGL5,FOG_CGL6,FOG_CGL7,FOG_CGL8,CasimirFOG,anu,tino,cqg,FOGST}.
Due to the large amount of possible models, an important issue is to select viable ones by experiments and observations. Therefore, the new born {\it multimessenger astronomy} is giving important  constraints to admit or exclude gravitational theories (see e.g. \cite{lombrisier,DeLaurentis}). However, also fine experiments can be conceived and realized in order to fix possible deviations and extensions with respect to GR. They can involve  space-based setups like satellites and precise electromagnetic measurements \cite{LucaPhot}.

We constrain the sizes of new gravitational forces (inferred in ETG or other scenarios) by making use the data coming from the precession of Planets. For this purpose, we follow the paper by Adkins and McDonnell \cite{mcdell} (see also \cite{FengXu,Chashcine}), where it is  calculated the precession of Keplerian orbits under the influence of arbitrary central-force perturbations. In the limit of nearly circular orbits, the perturbed orbit equation takes the form ($u=1/r$)
\begin{equation}
\frac{d^2 u}{d \varphi^2} + u = \frac{G M}{h^2} - \frac{g(u)}{h^2}
\end{equation}
where $g(u) = r^2 \frac{F(r)}{m} \vert_{r=1/u}$ ($\frac{F}{m}=-\nabla V$) and $h^2=G M a$.  $g(u)=0$ corresponds to the unperturbed solution.
We refer to the corrections to the Planets precession induced by the Yukawa-like potential, $V_Y(r)=\alpha \frac{e^{-\beta r}}{r}$, and power law (PL) potentials, $V_{PL}(r) = \alpha_n r^n$. In GR, the first post-Newtonian correction is a perturbing potential given by  $V(r)\Big|_{\text{GR}} = - \frac{G M h^2}{c^2 r^3}$, which corresponds the precession $\Delta \theta_p\Big|_{\text{GR}} = \frac{6 \pi G M}{c^2 a}$. This gives the well known $43$ arcsec per century when applied to the orbit of Mercury.
The correction to the Planet precessions induced by a generic perturbing force $F(z)$ and perturbing potential $V(z)$ is \cite{mcdell}
\begin{eqnarray} \label{general_result_2}
\Delta \theta_p &=& -\frac{2 a^2}{G M  \epsilon} \int_{-1}^1 \frac{dz \, z}{\sqrt{1-z^2}} \frac{F(z)}{(1+\epsilon z)^2} \\
    &=& -\frac{2 a}{G M  \epsilon^2} \int_{-1}^1 \frac{dz \, z}{\sqrt{1-z^2}} \frac{d V(z)}{dz}\,, \label{general_result_3}
\end{eqnarray}
where, for the sake of  convenience, the correction $\Delta \theta_p$ is written in terms of the dimensionless integration variable $z$ with a fixed range, while
$\epsilon$ is the eccentricity ($\epsilon < 1$). The perturbing force $F(z)$ and $V(z)$ are evaluated at radius $r=a/(1+\epsilon z)$. In the following we refer to the Yukawa-like and power-law potentials following from different gravitational theories of gravity.


\begin{itemize}

\item {\bf The Yukawa force} -  The Yukawa potential (as a correction to the Newtonian potential $V_N=GM/r$) is of the form \cite{Yukawa35,Nieto91}
\begin{equation}\label{Yukawapot}
V_Y(r) = \alpha \frac{e^{-r/\lambda}}{r}\equiv \alpha \frac{e^{-\beta r}}{r}
\end{equation}
where $\alpha$ and $\lambda\equiv 1/\beta$ are the strength and the range of the interaction, respectively. As we will see, such a potential occurs in several modified theories of gravity. The precession due to a Yukawa perturbation depends on two parameters: a range parameter $\kappa=a/\lambda=\beta a$ and the eccentricity $\epsilon$, i.e. $\Delta \theta_p(\kappa,\epsilon)$, where $a$  is the semi-major axis.  According to \cite{mcdell}, the correction to the precession is of the integral form
\begin{equation}\label{DeltaYukawa}
\Delta \theta_p(\kappa,\epsilon) = -\frac{2 \alpha}{G M \epsilon}\, I_{\epsilon, \beta}\,,
\end{equation}
where
\begin{equation}\label{Iepsilonk}
I_{\epsilon, \beta}\equiv  \int_{-1}^1 \frac{dz \, z}{\sqrt{1-z^2}}
\left ( 1+ \frac{\kappa}{1+\epsilon z} \right ) e^{-{\frac{\kappa}{1+\epsilon z}}} \, .
\end{equation}
The behavior of the integral (\ref{Iepsilonk}) is represented in Fig. \ref{fig:foobar} for several Planets.
\item {\bf Power Law potential} - The power law potential is of the form
 \begin{equation}\label{DeltaPowerla}
 V_{PL}(r) = \alpha_{q}\, r^{q}\,,
 \end{equation}
where the parameter $q$ assume arbitrary values. The precession (\ref{general_result_2}) can be exactly integrated, and leads to \cite{mcdell}
\begin{equation} \label{precesspowerlaw}
\Delta \theta_p(q) = \frac{-\pi \alpha_q}{G M } a^{q+1} \sqrt{1-\epsilon^2} \chi_q(\epsilon)\,,
\end{equation}
where $\chi_q(\epsilon)$ is written in terms of the Hypergeometric function
\begin{equation}
\chi_q(\epsilon) = q (q+1) \, {_2F_1} \left (\frac{1}{2}-\frac{q}{2},1-\frac{q}{2}\, ;2\, ;\epsilon^2 \right ) \, .
\end{equation}
\end{itemize}

These potentials occur in ETG and in Non-Commutative Spectral Geometry (the Yukawa-like potential), and Quintessence field surrounding a massive gravitational source (the Power Law potential). We shall infer the corrections to periastron advance for Solar Planets, referring in particular to Mercury, Mars, Jupiter and Saturn, as well as to S2 star orbiting around Sagittarius $A^*$.

The paper is organized as follows. In Section II we study the weak field limit of Scalar-Tensor Fourth Order Gravity models, and the effects of these corrections to the periastron advance of Planets. As a particular case, we consider the Non-Commutative Spectral Gravity. In Section III we study the effect of the Quintessence field present around a Schwarzschild Black Hole, and derive a lower bound on the adiabatic index of equation of state.
Finally, conclusions are drawn in last Section.





\section{Scalar-Tensor-Fourth-Order Gravity}\label{ETG1}

The action for ETG is given by (see for example \cite{Felix})

\begin{eqnarray}\label{FOGaction}
\mathcal{S}\,=\,\int d^{4}x\sqrt{-g}\biggl[f(R,R_{\mu\nu}R^{\mu\nu},\phi)+\omega(\phi)\phi_{;\alpha}\phi^{;\alpha}+\mathcal{X}\mathcal{L}_m\biggr],
\end{eqnarray}
where $f$ is a generic function of the invariant $R$ (the Ricci scalar ), the invariant $R_{\mu\nu}R^{\mu\nu}\,=\,Y$ ($R_{\mu\nu}$ is the Ricci tensor), the scalar field $\phi$, $g$ is the determinant of metric tensor $g_{\mu\nu}$ and $\mathcal{X}\,=\,8\pi G$. The Lagrangian density $\mathcal{L}_m$ is the minimally coupled ordinary matter Lagrangian density, $\omega(\phi)$ is a generic function of the scalar field.

The field equations obtained by varying the action (\ref{FOGaction}) with respect to $g_{\mu\nu}$ and $\phi$, In the metric approach, are\footnote{We use, for the Ricci tensor, the convention
$R_{\mu\nu}={R^\sigma}_{\mu\sigma\nu}$, whilst for the Riemann
tensor we define ${R^\alpha}_{\beta\mu\nu}=\Gamma^\alpha_{\beta\nu,\mu}+\cdots$. The
affine connections are the  Christoffel symbols of the metric, namely
$\Gamma^\mu_{\alpha\beta}=\frac{1}{2}g^{\mu\sigma}(g_{\alpha\sigma,\beta}+g_{\beta\sigma,\alpha}
-g_{\alpha\beta,\sigma})$, and we adopt the signature is $(-,+,+,+)$.}:
\begin{eqnarray}
\label{fieldequationFOG}
&&f_RR_{\mu\nu}-\frac{f+\omega(\phi)\phi_{;\alpha}\phi^{;\alpha}}{2}g_{\mu\nu}-f_{R;\mu\nu}+g_{\mu\nu}\Box
f_R+2f_Y{R_\mu}^\alpha
R_{\alpha\nu}+
\\\nonumber\\\nonumber
 &&-2[f_Y{R^\alpha}_{(\mu}]_{;\nu)\alpha}+\Box[f_YR_{\mu\nu}]+[f_YR_{\alpha\beta}]^{;\alpha\beta}g_{\mu\nu}+\omega(\phi)\phi_{;\mu}\phi_{;\nu}\,=\,
\mathcal{X}\,T_{\mu\nu}\,,\nonumber\\
\nonumber\\
\label{FE_SF}
&&2\omega(\phi)\Box\phi+\omega_\phi(\phi)\phi_{;\alpha}\phi^{;\alpha}-f_\phi\,=\,0~.
\end{eqnarray}
where:
\[f_R\,=\,\frac{\partial f}{\partial R}, \,\,\,\,\,\,f_Y\,=\,\frac{\partial f}{\partial Y}, \,\,\,\,\,\,\omega_\phi\,=\,\frac{d\omega}{d\phi}\,, \,\,\,\,  f_\phi\,=\,\frac{d f}{d\phi}\,, \]
and $T_{\mu\nu}\,=\,-\frac{1}{\sqrt{-g}}\frac{\delta(\sqrt{-g}\mathcal{L}_m)}{\delta
g^{\mu\nu}}$ is the the energy-momentum tensor of matter. We confine ourselves to the case in which the generic function $f$ can be expanded as follows
(notice that the all other possible contributions in $f$ are negligible \cite{PRD1,mio2,FOG_CGL,FOGST})
\begin{eqnarray}
\label{LimitFramework2}
f(R,R_{\alpha\beta}R^{\alpha\beta},\phi)\,=\,&&f_R(0,0,\phi^{(0)})\,R+\frac{f_{RR}(0,0,\phi^{(0)})}{2}\,R^2+
\frac{f_{\phi\phi}(0,0,\phi^{(0)})}{2}(\phi-\phi^{(0)})^2\nonumber\\\\\nonumber&&+f_{R\phi}(0,0,\phi^{(0)})R\,\phi+
f_Y(0,0,\phi^{(0)})R_{\alpha\beta}R^{\alpha\beta}~.
\end{eqnarray}
To study the weak-field approximation, we perturb Eqs.~(\ref{fieldequationFOG}) and (\ref{FE_SF}) in a Minkowski background $\eta_{\mu\nu}$~\cite{PRD1}, i.e. we look for perturbed solutions of the form
\begin{eqnarray}\label{MeticCart}
&&g_{\mu\nu}\,\simeq\,
\begin{pmatrix}
-1-2\Phi & 2\textbf{A} \\
2\textbf{A} & (1-2\Psi)\delta_{ij}\end{pmatrix}~.
\end{eqnarray}
and
\begin{eqnarray}
\nonumber
&&\phi\,\sim\,\phi^{(0)}+\phi^{(2)}+\dots\,=\,\phi^{(0)}+\varphi.
\end{eqnarray}
For matter described as a perfect fluid, hence $T_{00}\,=\,\rho$ and $T_{ij}\,=\,0$, one gets that, for a ball-like source with radius ${\cal R}$, the gravitational potentials $\{\Phi, \Psi, A_i\}$ and the scalar field $\varphi$ take the form ($c=1$) \cite{FOG_CGL,CasimirFOG,FOGST}
%
\begin{eqnarray}\label{ST_FOG_FE_NL_sol_ball}
\Phi(\mathbf{x}) &=& -\frac{GM}{|\mathbf{x}|}\big[1+\zeta(|\mathbf{x}|)\big]\,, \label{PhiFOG}\\
\zeta(|\mathbf{x}|)&\equiv& g(\xi,\eta)\,F(m_+ {\cal R})\,e^{-m_+|\mathbf{x}|}+\Big[\frac{1}{3}-g(\xi,\eta)\Big]\,F(m_- {\cal R})\,e^{-m_-|\mathbf{x}|}-
\frac{4\,F(m_Y {\cal R})}{3}\,e^{-m_Y|\mathbf{x}|}\,, \label{zetaFOG}\\
\Psi(\mathbf{x}) &=& -\frac{GM}{|\mathbf{x}|}\big[1-\psi(|\mathbf{x}|)\big]\,, \label{PsiFOG} \\
\psi(|\mathbf{x}|) &\equiv & g(\xi,\eta)\,F(m_+ {\cal R})\,e^{-m_+|\mathbf{x}|}+\Big[\frac{1}{3}-g(\xi,\eta)\Big]\,F(m_- {\cal R})\,e^{-m_-|\mathbf{x}|}+
\frac{2\,F(m_Y {\cal R})}{3}\,e^{-m_Y|\mathbf{x}|}\,, \label{psiFOG}  \\
\mathbf{A}(\mathbf{x}) &=& -\frac{2\,G\,\big[1-{\cal A}(|\mathbf{x}|)\big]}{|\mathbf{x}|^2}\,\mathbf{x}\times\mathbf{J}\,, \label{AFOG} \\
{\cal A}(|\mathbf{x}|) &\equiv & (1+m_Y|\mathbf{x}|)\,e^{-m_Y|\mathbf{x}|}\,, \label{calFOG} \\
\varphi(\textbf{x}) &=&\frac{GM}{|\textbf{x}|} \sqrt{\frac{\xi}{3}}\,\frac{2}{\omega_+-\omega_-} \biggl[F(m_+ {\cal R})\,e^{-m_+\,|\textbf{x}|}\,-F(m_- {\cal R})\,e^{-m_-\,|\textbf{x}|}\biggr]\,, \label{varphiFOG}
\end{eqnarray}
where $\mathbf{J}\,=\,2M\mathcal{R}^2\mathbf{\Omega}_0/5$ is the angular momentum of the ball, $f_R(0,0,\phi^{(0)})\,=\,1$, $\omega(\phi^{(0)})\,=\,1/2$, and
\begin{eqnarray}
g(\xi,\eta)\, &=& \,\frac{1-\eta^2+\xi+\sqrt{\eta^4+(\xi-1)^2-2\eta^2(\xi+1)}}{6\sqrt{\eta^4+(\xi-1)^2-2\eta^2(\xi+1)}}\,, \label{g-func} \\
F(m\,\mathcal{R})\,&=& \,3\frac{m\,\mathcal{R} \cosh m\,\mathcal{R}-\sinh m\,\mathcal{R}}{m^3\mathcal{R}^3}\,, \label{F-func} \\
\xi\,&=& \,3{f_{R\phi}(0,0,\phi^{(0)})}^2\,, \quad \eta\,=\,\frac{m_\phi}{m_R} \,, \label{m-functions} \\
m_\pm^2 &=& m_R^2\,\omega_\pm\,, \label{mpm-func}\\
\omega_\pm\,&=& \,\frac{1-\xi+\eta^2\pm\sqrt{(1-\xi+\eta^2)^2-4\eta^2}}{2}\,, \label{omega-func} \\
{m_R}^2\, &\doteq & \,-\frac{f_R(0,0,\phi^{(0)})}{3f_{RR}(0,0,\phi^{(0)})+2f_Y(0,0,\phi^{(0)})}\,, \label{mR-fucnt} \\
        {m_Y}^2\, &\doteq & \,\frac{f_R(0,0,\phi^{(0)})}{f_Y(0,0,\phi^{(0)})}\,, \quad {m_\phi}^2\,\doteq\,-\frac{f_{\phi\phi}(0,0,\phi^{(0)})}{2\omega(\phi^{(0)})}\,.
        \label{csi-func}
\end{eqnarray}

Some ETG models  studied in literature and reported in  Table \ref{table} (see \cite{FOG_CGL} for further details).

\begin{center}
\begin{table*}[t]
\caption{\label{table} We report  different cases of Extended Theories of Gravity including a scalar field and higher-order curvature terms. The free parameters  are given as effective masses with their  asymptotic behavior. Here, we assume  $f_R(0,\,0,\,\phi^{(0)})\,=\,1$,  $\omega(\phi^{(0)})\,=\,1/2$.}
{\small
\hfill{}
\begin{tabular}{|l|l|c|c|c|c|c|c|}
\hline
\multicolumn{1}{|c|}{\textbf{Case}}&\multicolumn{1}{|c|}{\textbf{ETG}}& \multicolumn{5}{|c|}{\textbf{Parameters}}\\
\cline{3-7}
& & $m^2_R$ & $m^2_Y$ &$m^2_\phi$&$m^2_+$&$m^2_-$
 \\
\hline
\hline
A&\tiny{$f(R)$ }&\tiny{$-\frac{f_{R}(0)}{3f_{RR}(0)}$}&$\infty$& 0 & $m^2_R$ & $\infty$ 
\\
\hline
\hline
B&\tiny{$f(R,R_{\alpha\beta}R^{\alpha\beta})$}&\tiny{$-\frac{f(0)}{3f_{RR}(0)+2f_Y(0)}$}&\tiny{$\frac{f_{R}(0)}{f_Y(0)}$}& 0 & $m^2_R$ & $\infty$
\\
\hline
\hline
C&\tiny{$f(R,\phi)+\omega(\phi)\phi_{;\alpha}\phi^{;\alpha}$}&\tiny{$-\frac{f_{R}(0)}{3f_{RR}(0)}$}& $\infty$ &\tiny{$-\frac{f_{\phi\phi}(0)}{2\omega(\phi^{(0)})}$}&\tiny{$m^2_R w_{+}$}&\tiny{$m^2_R w_{-}$}
\\
\hline
\hline
D&\tiny{$f(R,R_{\alpha\beta}R^{\alpha\beta},\phi)+\omega(\phi)\phi_{;\alpha}\phi^{;\alpha}$} &\tiny{$-\frac{f(0)}{3f_{RR}(0)+2f_Y(0)}$}&\tiny{$\frac{f_{R}(0)}{f_Y(0)}$}&\tiny{$-\frac{f_{\phi\phi}(0)}{2\omega(\phi^{(0)})}$}&\tiny{$m^2_R w_{+}$}&\tiny{$m^2_R w_{-}$}
\\
\hline
\end{tabular}}
\hfill{}
\end{table*}
\end{center}

\subsection{Planet precession in fourth order gravity}

\begin{figure}
    \centering
    \subfigure[]{\includegraphics[width=0.4\textwidth]{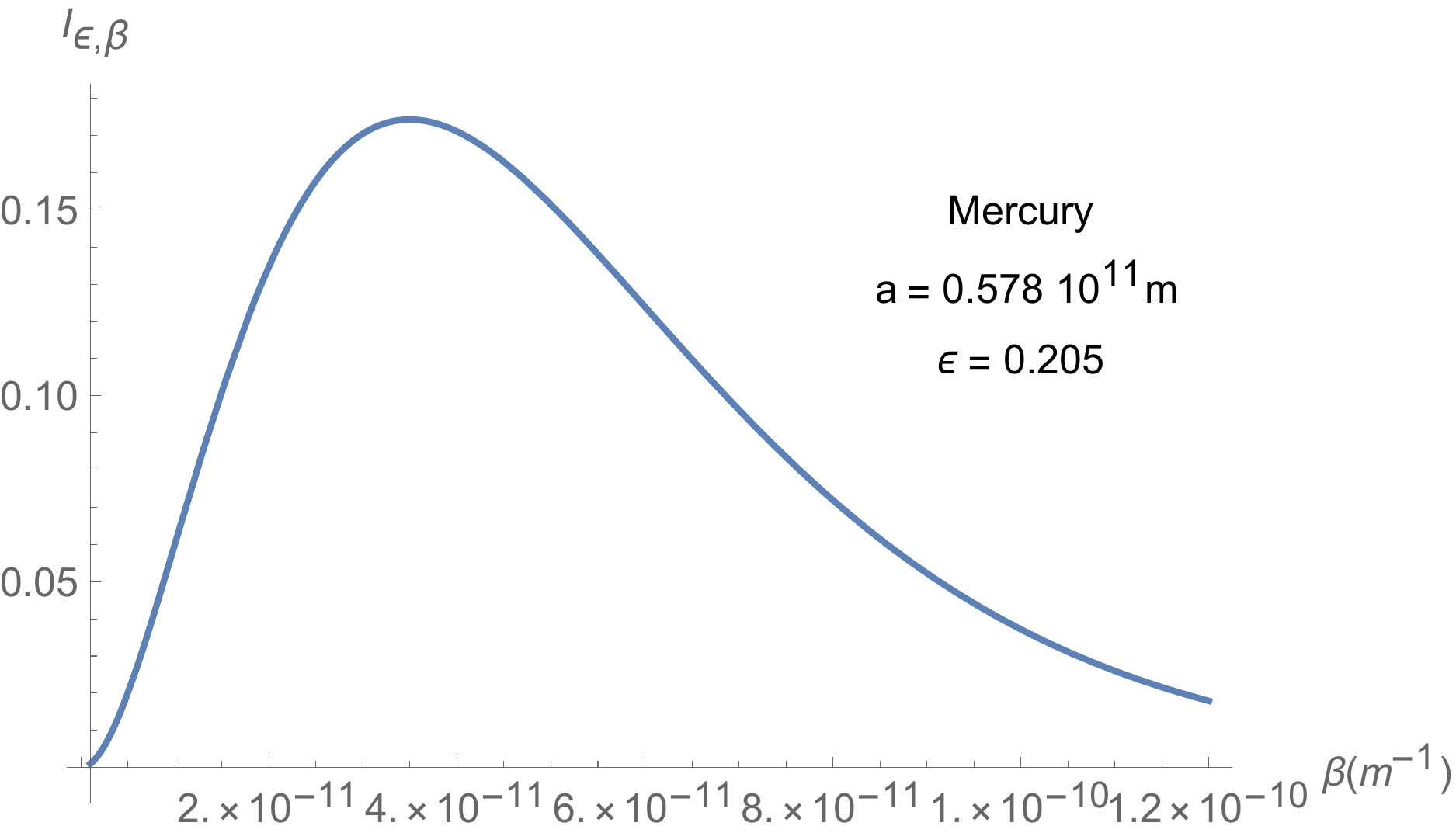}}
    \subfigure[]{\includegraphics[width=0.4\textwidth]{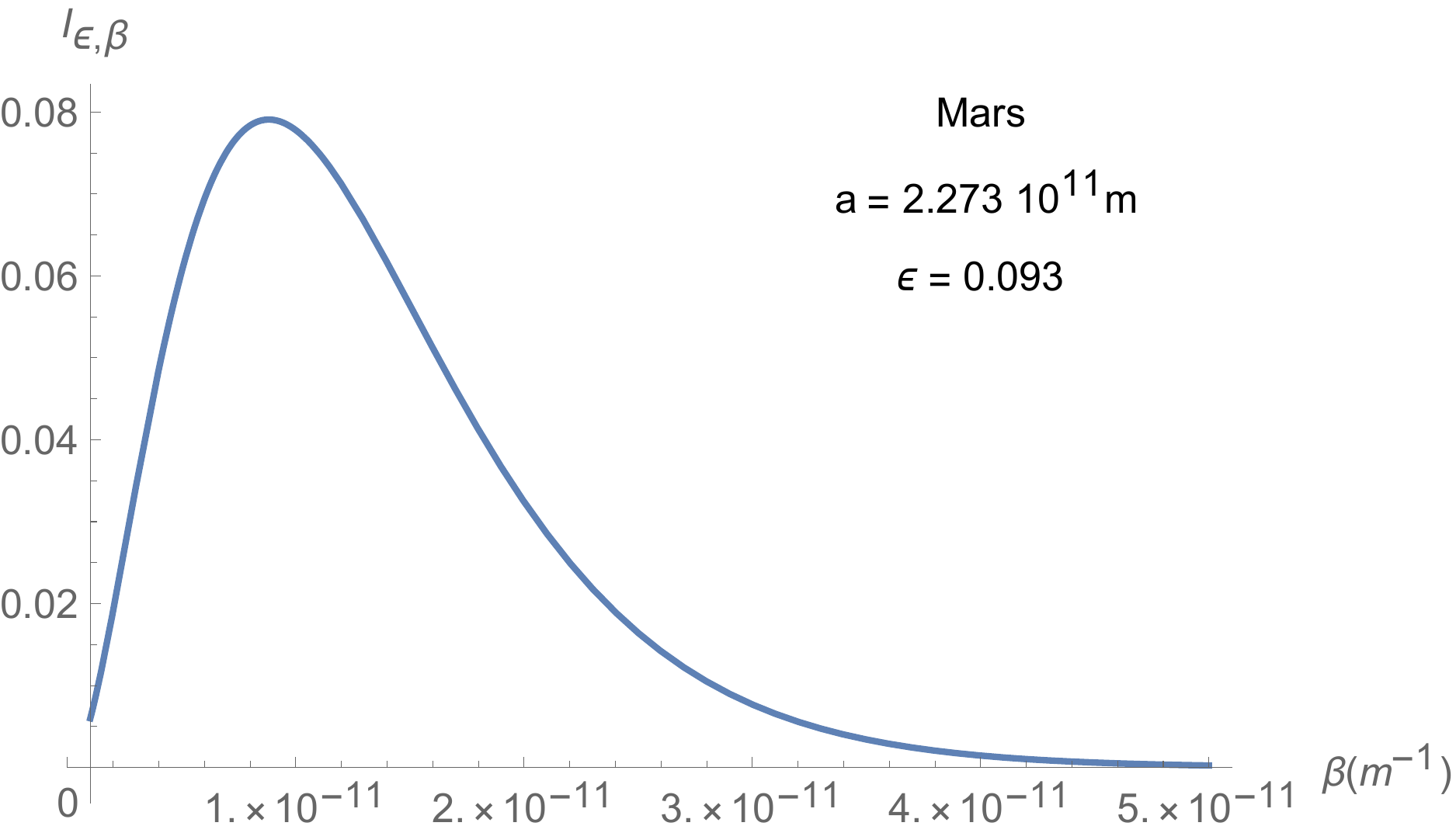}}
    \subfigure[]{\includegraphics[width=0.4\textwidth]{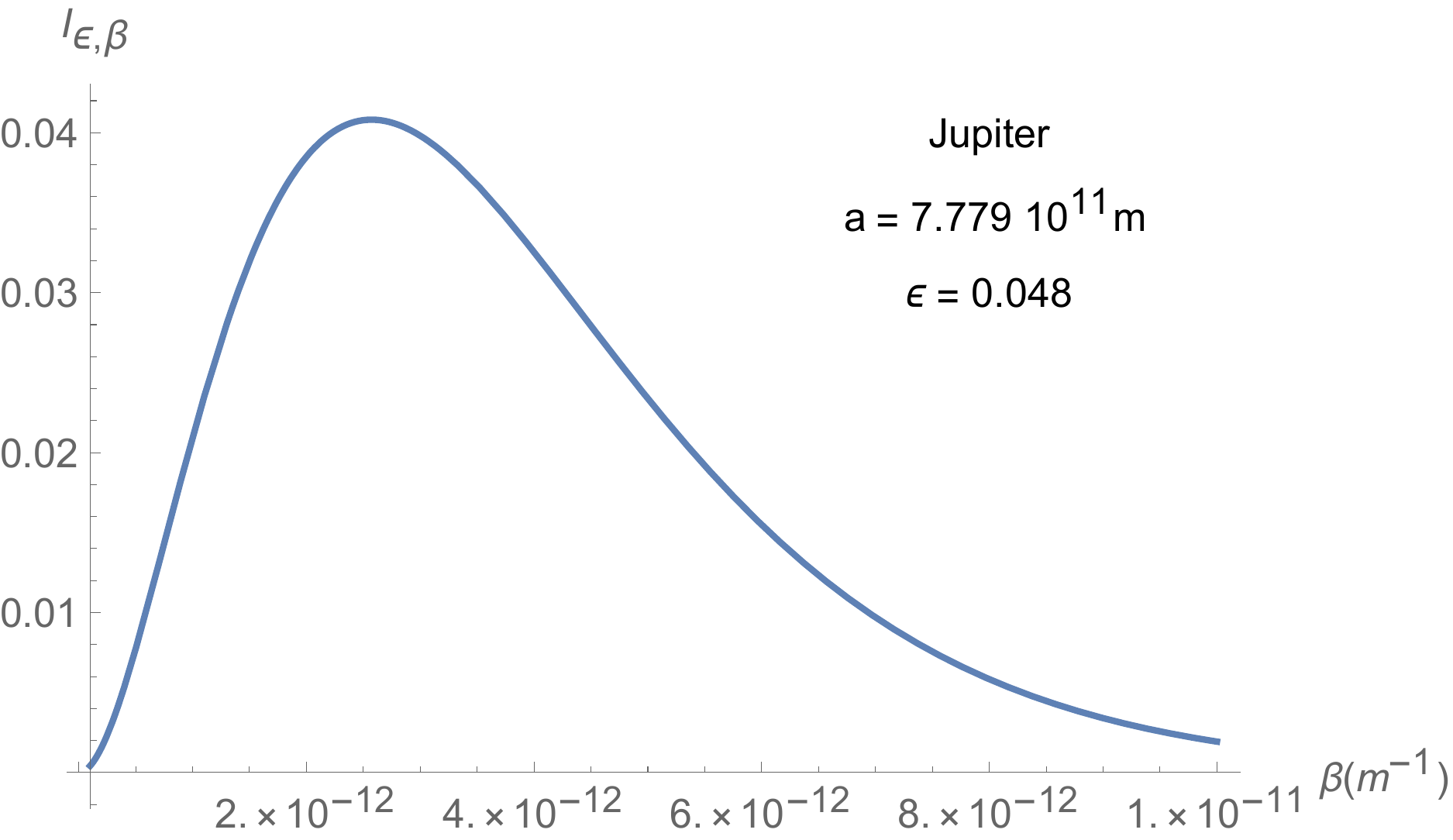}}
    \subfigure[]{\includegraphics[width=0.4\textwidth]{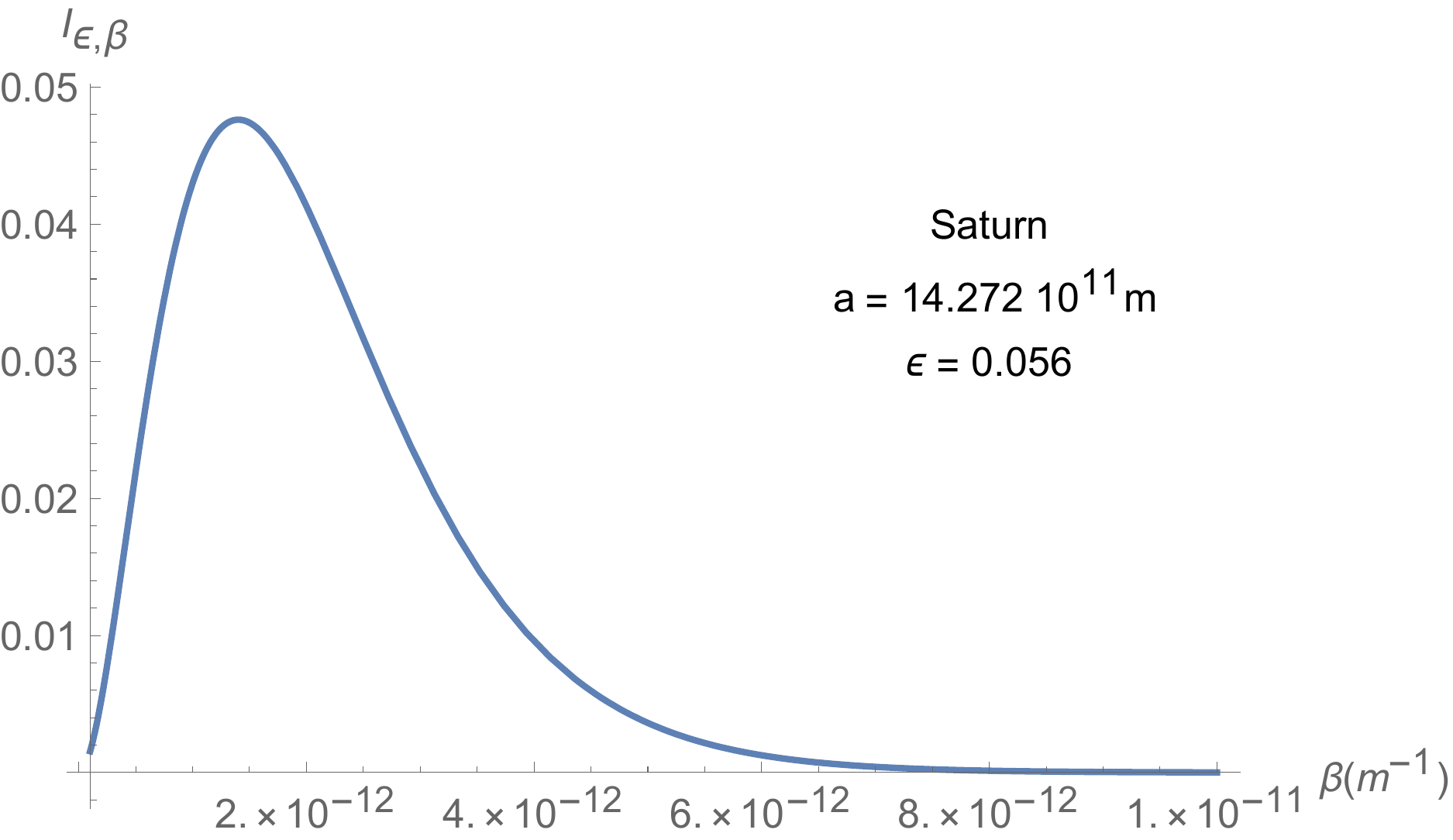}}
    \caption{(a) $I_{\epsilon, \beta}$ vs $\beta$ for Mercury. (b) $I_{\epsilon, \beta}$ vs $\beta$ for Mars. (c) $I_{\epsilon, \beta}$ vs $\beta$ for Jupiter. (d) $I_{\epsilon, \beta}$ vs $\beta$ for Saturn.}
    \label{fig:foobar}
\end{figure}

\begin{figure}
    \centering
    \subfigure[]{\includegraphics[width=0.45\textwidth]{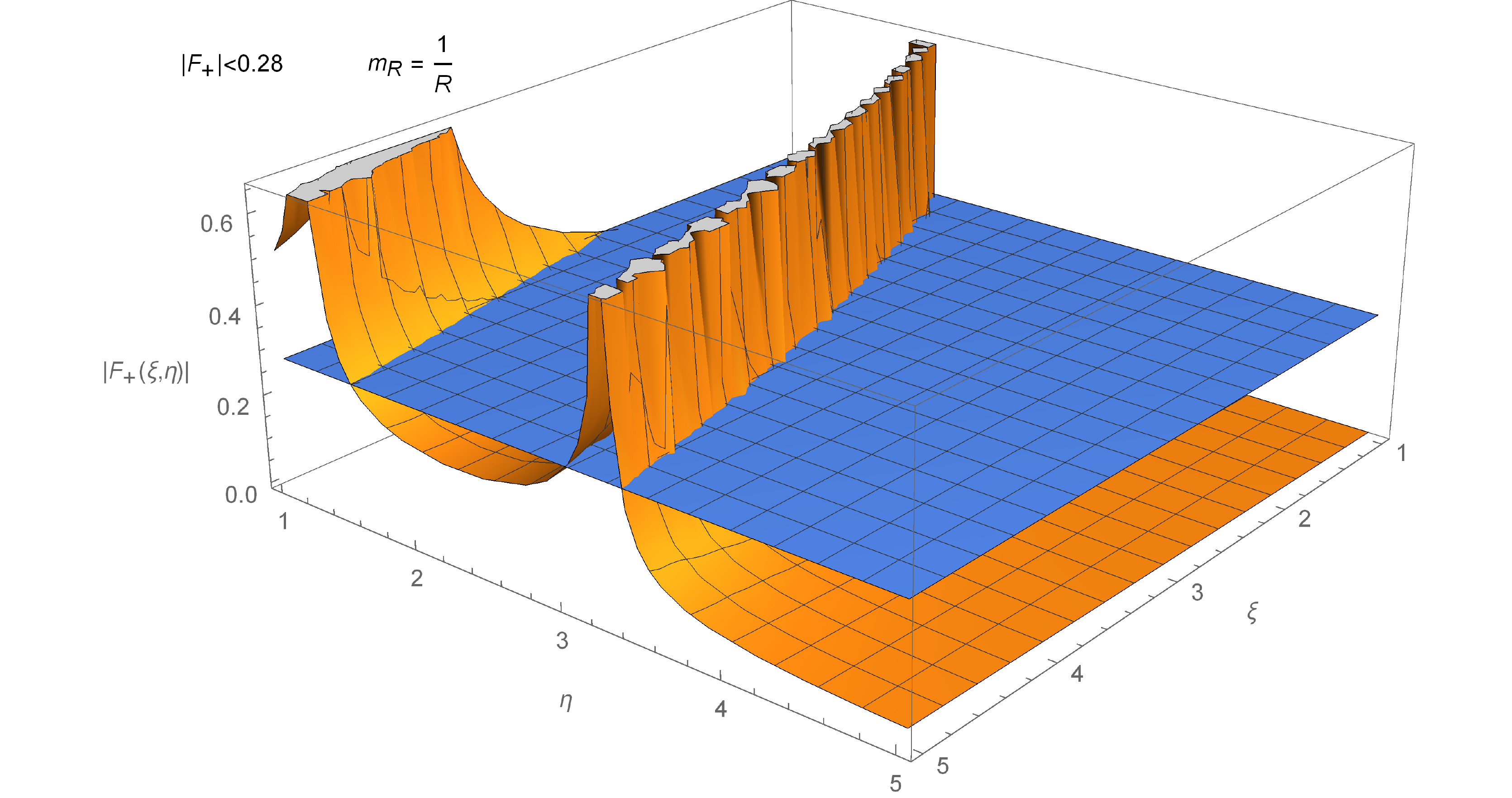}}
    \subfigure[]{\includegraphics[width=0.45\textwidth]{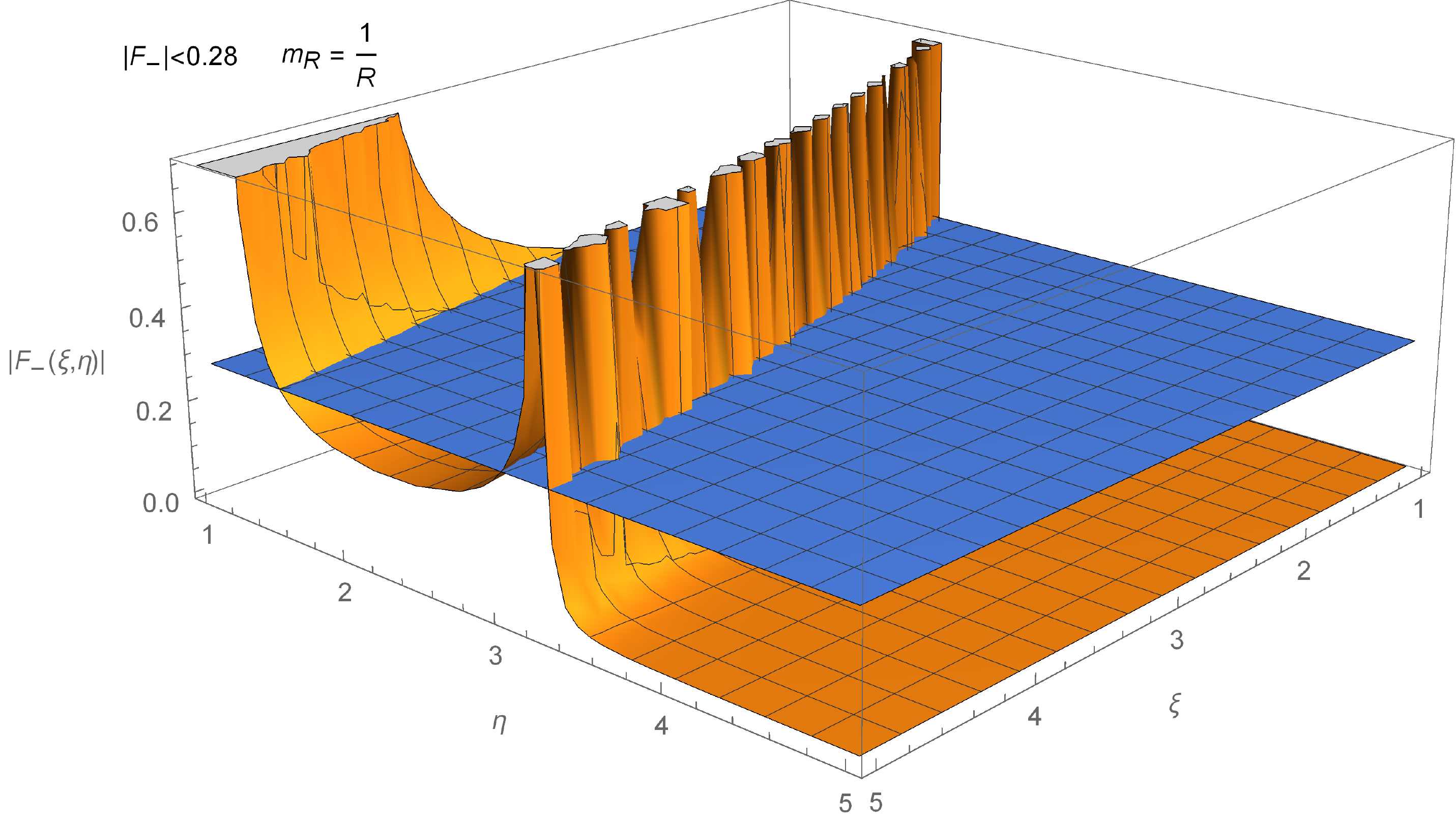}}
    \caption{(a) $F_+$ vs $\{\xi, \eta\}$ for Mercury ($|F_+| \lesssim 0.28$), with $m_R=\frac{1}{{\cal R}}$. (b) $F_-$ vs $\{\xi, \eta\}$ for Mercury ($|F_-| \lesssim 0.28$) with $m_R=\frac{1}{{\cal R}}$.}
    \label{FPlusMinus3D}
\end{figure}

In this Section we study the periastron shift of the orbital period of objects, both astrophysical and Solar System, in four order gravity (FOG). As we have seen, the FOG field equations lead to a gravitational potential of the Yukawa-like form ($r=|{\bf x}|$)
 \begin{equation}\label{VFOG}
 V(r)=\frac{GM}{r} \left(1+\sum_{i=\pm, Y} F_i e^{-\beta_i r}\right)\,,
 \end{equation}
where $F_i$ and $\beta$ are the strength and range of the interaction corresponding to each mode $i=+, -, Y$. Comparing (\ref{VFOG}) with (\ref{Yukawapot}),
it follows the correspondence (referring to (\ref{PhiFOG}) and (\ref{zetaFOG}))
\begin{equation}\label{alphaFOG}
 \alpha \to GM F_i\,, \quad \beta \to \beta_i\,, \quad i=\pm, Y\,.
\end{equation}
with
\begin{equation}\label{FparameterFOG}
 F_+ =g(\xi,\eta)\,F(m_+ {\cal R})\,, \quad F_- = \Big[\frac{1}{3}-g(\xi,\eta)\Big]\,F(m_- {\cal R})\,, \quad
 F_Y=- \frac{4}{3}\,F(m_Y {\cal R})\,,
\end{equation}
\begin{equation}\label{betaFOG}
 \beta_\pm = m_R \sqrt{\omega_\pm} \,, \qquad \beta_Y = m_Y\,.
\end{equation}
We impose that the periastron shift $\Delta \theta_p(\kappa,\epsilon) = -\frac{2 \alpha}{G M \epsilon}\, I_{\epsilon, \beta}$ given by (\ref{DeltaYukawa}), where $I_{\epsilon, \beta}$ is defined in (\ref{Iepsilonk}), is lesser than the error $\eta$. Fixing $I_{\epsilon, \beta}$ to the maximum values, one gets the bounds on the parameters $F_i$:
 \begin{equation}\label{Fibound}
 |\Delta \theta_p(\kappa,\epsilon)|\lesssim \eta \quad \to \quad  |F_i|\lesssim \frac{\eta \epsilon}{2I_{\epsilon, \beta_i}}\,, \,\, i=\pm, Y\,.
 \end{equation}
In Fig. \ref{fig:foobar} are plotted the function $I_{\epsilon, \beta}$ for the Mercury, Mars, Jupiter and Saturn planets. In Table III  are reported the
corresponding bounds on $F_i$. As an illustrative example, we plot $|F_\pm (\xi, \eta)|$ in Fig. \ref{FPlusMinus3D}, for $m_R={\cal R}^{-1}$. The available values of the parameters $\{\xi, \eta\}$ allow to fix the masses, via Eqs. (\ref{g-func}) ,(\ref{F-func}), (\ref{m-functions}), (\ref{mR-fucnt}), of extra modes arising in Scalar Tensor Fourth Order Gravity. The analysis of Yukawa gravitational potential for $f(R)$ has been carried out in \cite{DeMartino}.

\begin{table}
\begin{center}
\caption{Values of periastron advance for the first six planets of the Solar System. In the table we present the values of the eccentricity $\epsilon$, semi-major axis $a$ in meters, the orbital period $P$ in years, the periastron advance predicted in General Relativity (GR).}\label{TabPlanets}
\begin{tabular}{|c||ccccc|}
	\hline
	Planet & $\qquad\epsilon\qquad$ & $\quad a \, (10^{11}\, m)\quad$ & $P\, (yrs)$ & $\quad \Delta\phi_{GR} \, (''/century)\quad $ & $\Delta\phi_{obs}$ \\
	\hline \hline
	Mercury &  $0.205$ & $0.578$  &  $0.24$ & $43.125$
	& $42.989   \pm 0.500$ \\
	\hline
	Venus & $0.007$  & $1.077$ & $0.62$  &  $8.62$ &  $8.000\pm 5.000$   \\
	\hline
	Earth & $0.017$  &  $1.496$ &  $1.00$ & $3.87$  &    $5.000\pm 1.000$  \\
	\hline
	Mars  & $0.093$  &  $2.273$ & $1.88$  & $1.36$  &   $1.362\pm 0.0005$   \\
	\hline
	Jupiter & $0.048$  &  $7.779$ & $11.86$   &  $0.0628$  & $0.070 \pm 0.004$  \\
	\hline
	Saturn &  $0.056$ &   $14.272$ &  $29.46$  &   $0.0138$ &   $0.014 \pm 0.002$ \\
	\hline
\end{tabular}
\end{center}
\end{table}

\begin{table}
\begin{center}
\caption{Bounds on $F_i$, $i=\pm, Y$ obtained from (\ref{Fibound})  using the values of periastron advance for planets of the Solar System. }
\begin{tabular}{|c||c|c|c|c|}
	\hline
	Planet & $|\eta|$ & $I_{\epsilon, \beta}^{\text{max}}$ &  $\beta_i^{\text{max}}\simeq $ & $|F_i|\lesssim$ \\
	\hline \hline
Mercury & 0.5 & 0.18 & $4\times 10^{-11}m^{-1}$ & 0.28\\
Mars & $5 \times 10^{-4}$ & 0.08 & $1.1 \times 10^{-11}m^{-1}$ & $2.9 \times 10^{-4}$ \\
Jupiter & $4\times 10^{-3}$ & 0.04 & $2.5 \times 10^{-12}m^{-1}$ & $2.4 \times 10^{-3}$ \\
Saturn & $2\times 10^{-3}$ & 0.05 & $2\times 10^{-13}m^{-1}$ & $1.1 \times 10^{-3}$ \\
	\hline
\end{tabular}
\label{tableIIa}
\end{center}
\end{table}

\subsection{Non-commutative geometry}

As a special case of  scalar-tensor-fourth-order gravity theory that we want discuss is the Non-Commutative Spectral Geometry (NCSG) \cite{connes_1,connes_2}.
%
%
%
Among the various attempts to unify all interactions, including gravity, NCSG is one of the most interesting candidate \cite{ccm,ncg-book1,mairi2012}.
It proposes that the Standard Model (SM) fields and gravity are packaged into geometry and matter on a Kaluza-Klein noncommutative space.
In NCSG, geometry is composed by a two-sheeted space, made from the product of a four-dimensional compact Riemannian manifold ${\cal M}$ (with a fixed spin structure), describing the geometry of space-time, and a discrete noncommutative space ${\cal F}$, describing the internal space of the particle physics model.
The SM fields and gravity enter into matter and geometry on a noncommutative space which has the product form ${\cal M}\times {\cal F}$. Such a product space
is physically interpreted in the way that left- and right-handed fermions are placed on two different sheets with the Higgs fields being the gauge fields in the discrete dimensions (the Higgs can be seen as the difference (thickness) between the two sheets).
The choice of a two-sheet geometry has a deep physical meaning since such a structure accommodate the gauge symmetries of the SM, and incorporates the seeds of quantisation (see \cite{Sakellariadou:2011wv} for details, and references therein).

In the gravitational sector, to which we are interested in, the action includes the coupling between the Higgs field $\phi$ and the Ricci curvature scalar $R$ \cite{ccm}
\begin{equation}\label{eq:0}
S_{\rm grav} = \int \left(\frac{R}{2\kappa^2} + \alpha_0
C_{\mu\nu\rho\sigma}C^{\mu\nu\rho\sigma} + \tau_0 R R^\star - \xi_0 R|{\bf H}|^2 \right)
\sqrt{-g} {\rm d}^4 x\,,
\end{equation}
where $\kappa^2\equiv 8\pi G$, ${\bf H}=(\sqrt{af_0}/\pi)\phi$ is the Higgs field, with $a$ a parameter related to fermion and lepton masses and lepton mixing, while $C^{\mu\nu\rho\sigma}$ is the Weyl tensor (the square of the Weyl tensor can be expressed in terms of $R^2$ and $R_{\mu\nu}R^{\mu\nu}$: $C_{\mu\nu\rho\sigma}C^{\mu\nu\rho\sigma}\,=\,2R_{\mu\nu}R^{\mu\nu}-\frac{2}{3}R^2$) and $R {}^*R=\frac{1}{2}\epsilon^{\alpha\beta\gamma\delta}R_{\alpha\beta\sigma\rho}R_{\gamma\delta}^{\quad\sigma\rho}$ ($R^\star R^\star$ is the topological term related to the Euler characteristic). At unification scale (fixed  by the cutoff $\Lambda$), $\alpha_0=-3f_0/(10\pi^2)$.
The NCSG model offers a framework to study several topics
\cite{Nelson:2008uy,Sakellariadou:2012jz,Chamseddine:2005zk,Chamseddine:2008zj,cchiggs,Chamseddine:2013rta,stabile}). It is worth to note that the quadratic curvature terms in the action functional does not give rise to the emergence of negative  \cite{stelle}  energy massive graviton modes\footnote{The higher derivative terms that are quadratic in curvature lead to \cite{Donoghue:1994dn}
\begin{equation}
\int\left (\frac{1}{2\eta} C_{\mu\nu\rho\sigma}C^{\mu\nu\rho\sigma}
-\frac{\omega}{3\eta} R^2 +\frac{\theta}{\eta}E \right) \sqrt{-g}d^4x~;
\nonumber
\end{equation}
$E=R^\star R^\star$ denotes the topological term which is the integrand in the Euler characteristic $\int E\sqrt{-g}d^4x=\int R^\star R^\star \sqrt{-g}d^4x$
The running of the coefficients $\eta, \omega, \theta$ of the higher
derivative terms is determined by the renormalization group
equations~\cite{Donoghue:1994dn}.  The coefficient $\eta$ goes slowly to zero in the infrared limit, so that $1/\eta={\cal O}(1)$ up to
scales of the order of the size of the Universe. Note that $\eta(t)$ varies by at
most one order of magnitude between the Planck scale and infrared
energies.  All three coefficients $\eta(t), \omega(t), \theta(t)$ run to a
singularity at a very high energy scale ${\cal O}(10^{23}) {\rm GeV}$ (i.e., above the Planck scale).
To avoid low energy constraints, the coefficients of the quadratic curvature terms $R_{\mu\nu}R^{\mu\nu}$ and $R^2$ should not exceed
$10^{74}$~\cite{Donoghue:1994dn}, which is indeed the case for the running of these coefficients.} \cite{Donoghue:1994dn}.

\begin{figure}
    \centering
    \subfigure[]{\includegraphics[width=0.4\textwidth]{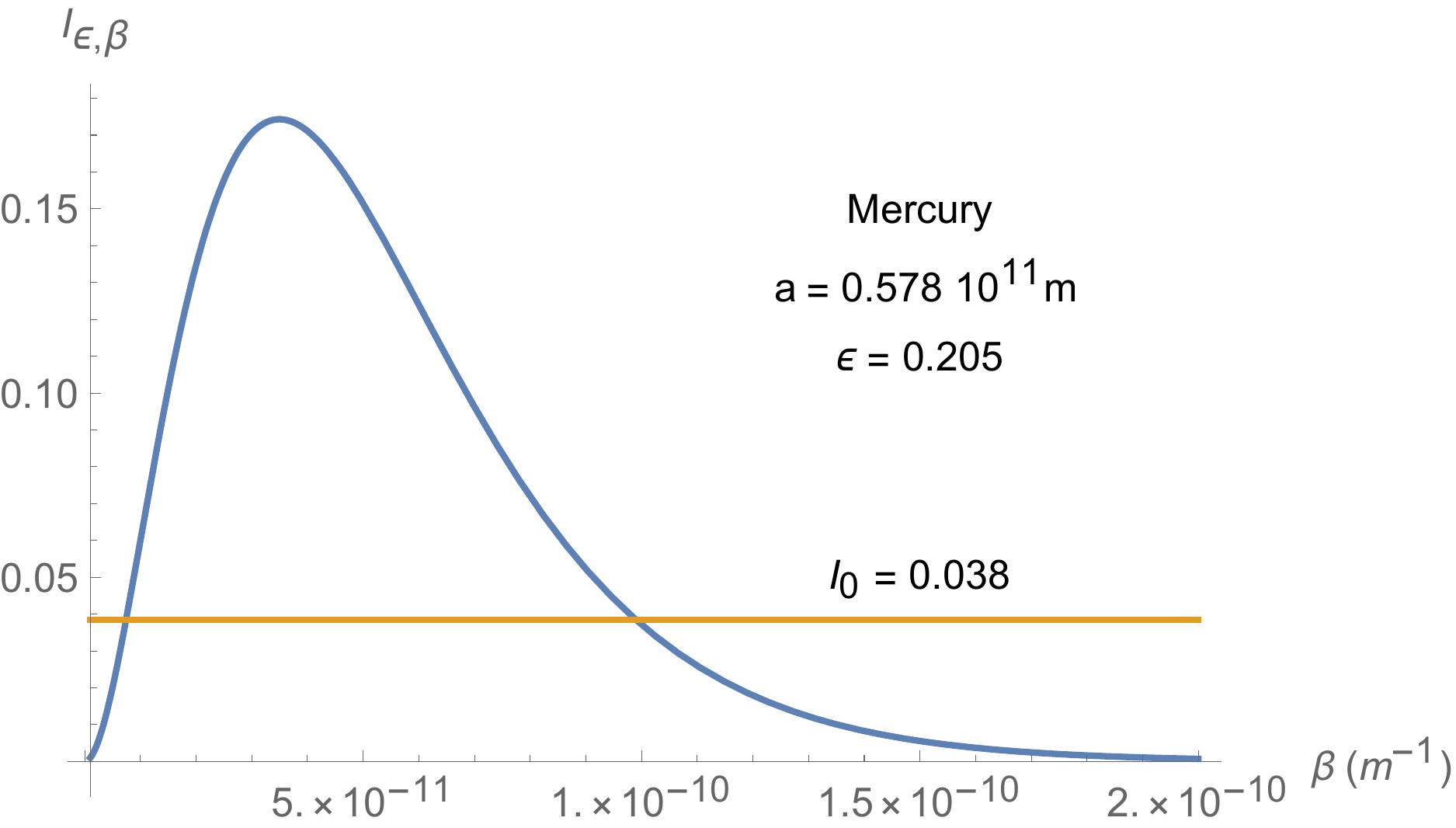}}
    \subfigure[]{\includegraphics[width=0.4\textwidth]{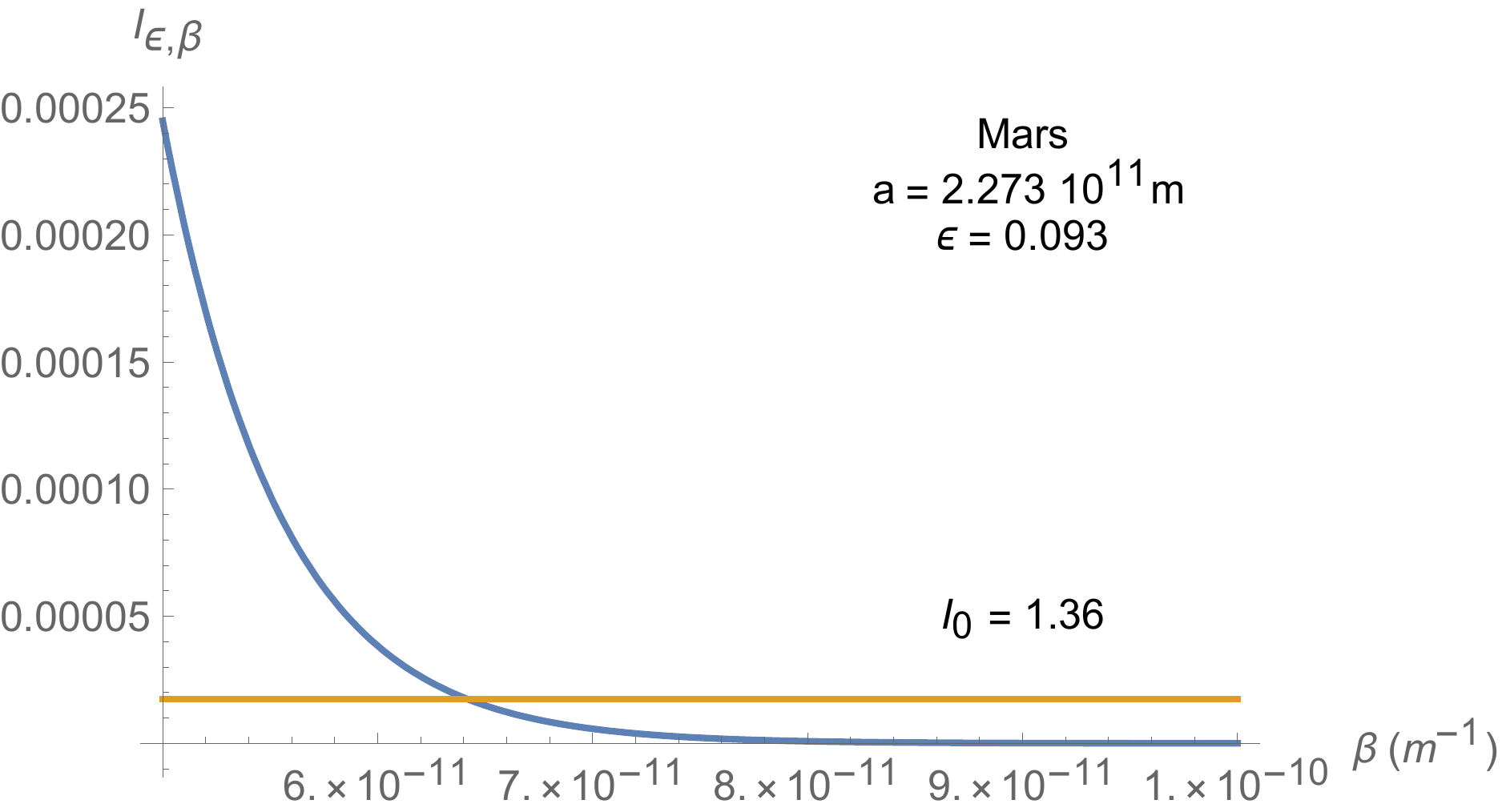}}
    \subfigure[]{\includegraphics[width=0.4\textwidth]{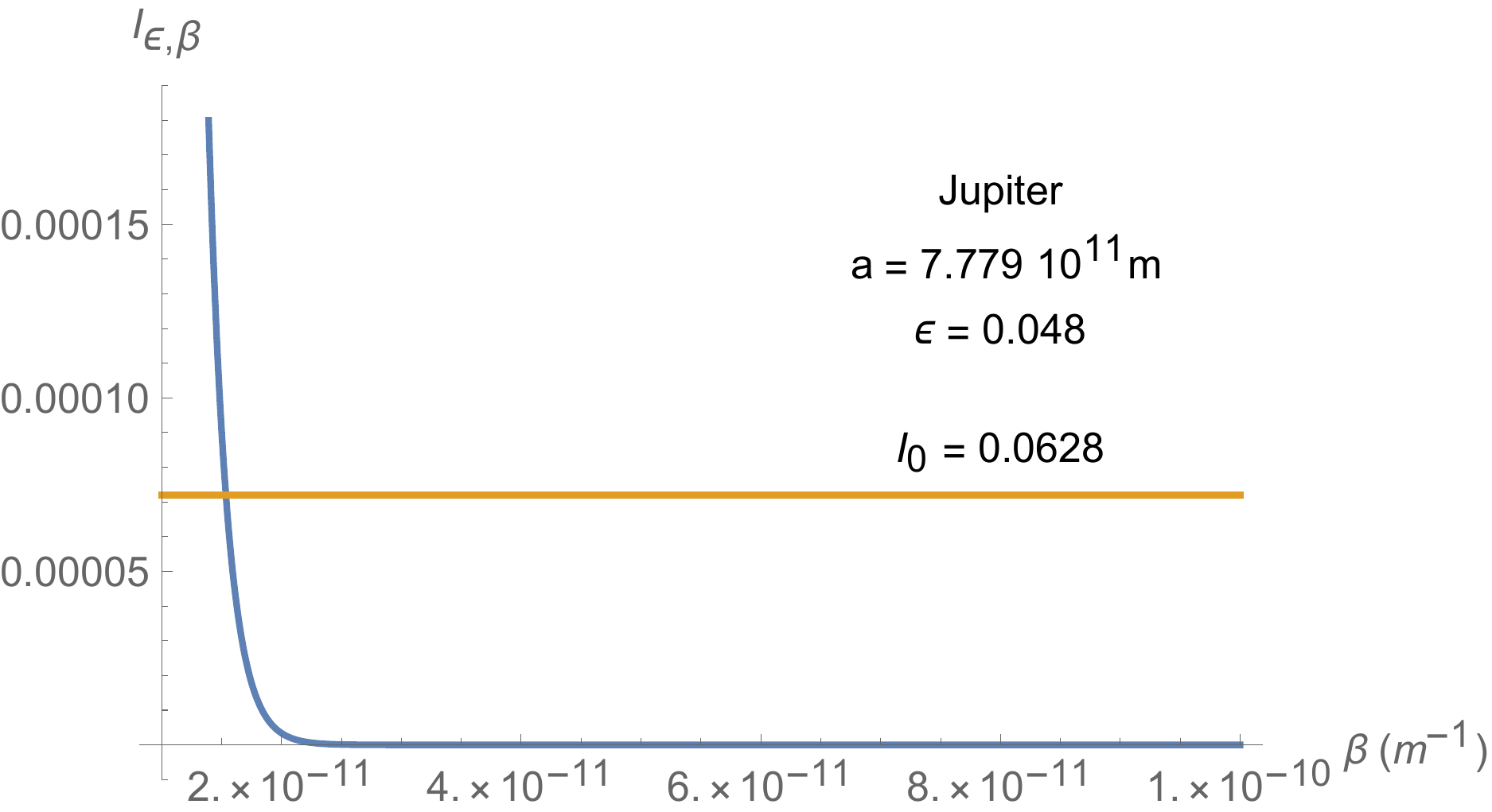}}
    \subfigure[]{\includegraphics[width=0.4\textwidth]{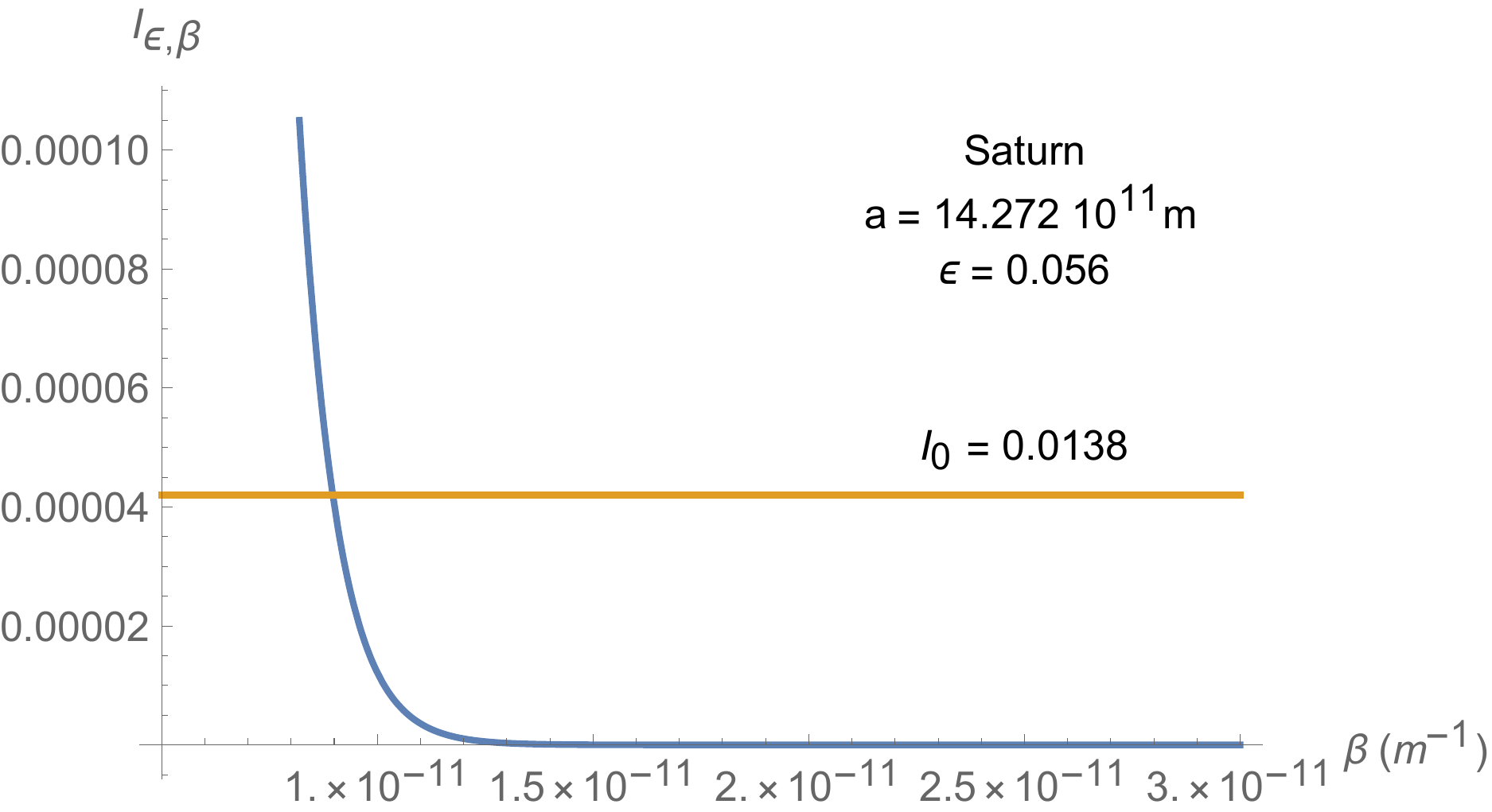}}
    \caption{(a) $I_{\epsilon, \beta}$ vs $\beta$ for Mercury. (b) $I_{\epsilon, \beta}$ vs $\beta$ for Mars. (c) $I_{\epsilon, \beta}$ vs $\beta$ for Jupiter. (d) $I_{\epsilon, \beta}$ vs $\beta$ for Saturn.}
    \label{fig:NCSGfoobar}
\end{figure}

The variation of the action (\ref{eq:0}) with respect to the metric tensor yields the NCSG equations of motion \cite{Nelson:2008uy}
\begin{equation}\label{2}
 G^{\mu\nu}+\frac{1}{\beta_{\small NCSG}^2}[2\nabla_\lambda \nabla\kappa
  C^{\mu\nu\lambda\kappa}+C^{\mu\lambda\nu\kappa}R_{\lambda\kappa}] = \kappa^2 T^{\mu\nu}_{({\rm matter})}\,,
\end{equation}
where $\beta_{\small NCSG}^{2}=\displaystyle{5\pi^2/(6\kappa^2f_0)}$.
In the weak field approximation, $g_{\mu\nu}=\eta_{\mu\nu}+\gamma_{\mu\nu}$, one gets \cite{stabile}.
 \begin{equation}\label{elementline}
    ds^2 = -(1+2\Phi)dt^2+ 2{\bf A}\cdot d{\bf x} dt+(1+2\Psi) d{\bf x}^2 \,,
 \end{equation}
with
\begin{eqnarray}\label{gamma001}
    \gamma_{00}&=&-2\Phi=\frac{2GM}{r}\left(1-\frac{4}{3}e^{-\beta_{\small{NCSG}} r}\right)\,,  \\
    \gamma_{0i}&=&\gamma_{i0}= A_i \\
    &=&-\frac{4G}{r^3}\left[1-(1+\beta_{\small{NCSG}} r)e^{-\beta_{\small{NCSG}} r}\right]({\bf r}\wedge {\bf J})_i\,, \nonumber \\
    \gamma_{ij} &=& 2\Psi \delta_{ij} = \frac{2GM}{r}\left(1+\frac{5}{9}e^{-\beta_{\small{NCSG}} r}\right)\delta_{ij}\,, \label{gamma0ij}
 \end{eqnarray}
The modifications induced by the NCSG action to the Newtonian potentials $\Phi$ (and $\Psi$), Eq.~(\ref{gamma001}), are similar to those induced by a Yukawa-like potential (\ref{Yukawapot}) (fifth-force \cite{fischbach}), with
 \begin{equation}
  \alpha=\frac{4}{3}GM\,, \quad \beta_{\small NCSG}=\beta
  \end{equation}
Following the previous section, Eqs. (\ref{DeltaYukawa}) (\ref{Fibound}), the periastron advance in NCSG for planets is given by
 \begin{equation}\label{betaNCSGbound}
 |\Delta \theta_p(\beta,\epsilon)|\lesssim \eta \quad \to \quad  |I_{\epsilon, \beta}|\lesssim I_0\,, \quad I_0\equiv \frac{3 \eta \epsilon}{8}\,,
 \end{equation}
where $I_{\epsilon, \beta}$ is defined in (\ref{Iepsilonk}). From Eq, (\ref{betaNCSGbound}) one infers  the bounds on $\beta$, or equivalently an upper bound on $\lambda$. Results are reported in Table \ref{tableNCG} (see also Fig. \ref{fig:NCSGfoobar}).
These results show that the bounds on $\beta$ improve several order of magnitude as compared with ones obtained using recent observations of pulsar timing,
$\beta \geq 7.55\times 10^{-13} {\rm m}^{-1}$ \cite{Nelson:2010rt, Nelson:2010ru}.
Bounds on the parameter $\beta$ have been obtained in different frameworks.  From Gravity Probe B experiment one gets $\beta>10^{-6} {\rm m}^{-1}$ \cite{stabile}. A more stringent constraint on $\beta$ can be obtained  from laboratory experiments designed to test the fifth force, that is, by constraining $\lambda$ through torsion balance experiments which implies to obtain a stronger lower bound on
$\beta$ (or equivalently an upper bound to the momentum $f_0$ in NCSG theory).
The test masses have a typical size of $\sim 10$mm and their separation is smaller than their size. As we have already mentioned
above, in NCSG one has $|\alpha| \sim {\cal O}(1)$, so that the tightest constraint on $\lambda=\beta^{-1}$ provided by E\"ot-Wash~\cite{eot} and
Irvine~\cite{irvine} experiments is~\cite{kapner} $\lambda \lesssim 10^{-4}\mbox{m}$, or equivalently $\beta \gtrsim 10^4 \mbox{m}^{-1}$.

\begin{table}
\begin{center}
\caption{Lower bounds on $\beta$ obtained from (\ref{betaNCSGbound})  using the values of periastron advance for planets of the Solar System. }
\begin{tabular}{|c||c|c|c|}
	\hline
	Planet & $\eta$ & $I_0 \equiv \frac{3\eta \epsilon}{8}$ &  $\beta (m^{-1}) > $ \\
	\hline \hline
Mercury & 0.5 & 0.038 & $1.0 \times 10^{-10}  $\\
Mars & $5 \times 10^{-4}$ & 1.36 & $7.8 \times 10^{-11}$ \\
Jupiter & $4\times 10^{-3}$ & 0.0628 & $2.1 \times 10^{-11}$ \\
Saturn & $2\times 10^{-3}$ & 0.0138 & $8.5 \times 10^{-12}$ \\
	\hline
\end{tabular}
\label{tableNCG}
\end{center}
\end{table}

\section{Quintessence - Dark energy}

\begin{figure}
    \centering
    \subfigure[]{\includegraphics[width=0.4\textwidth]{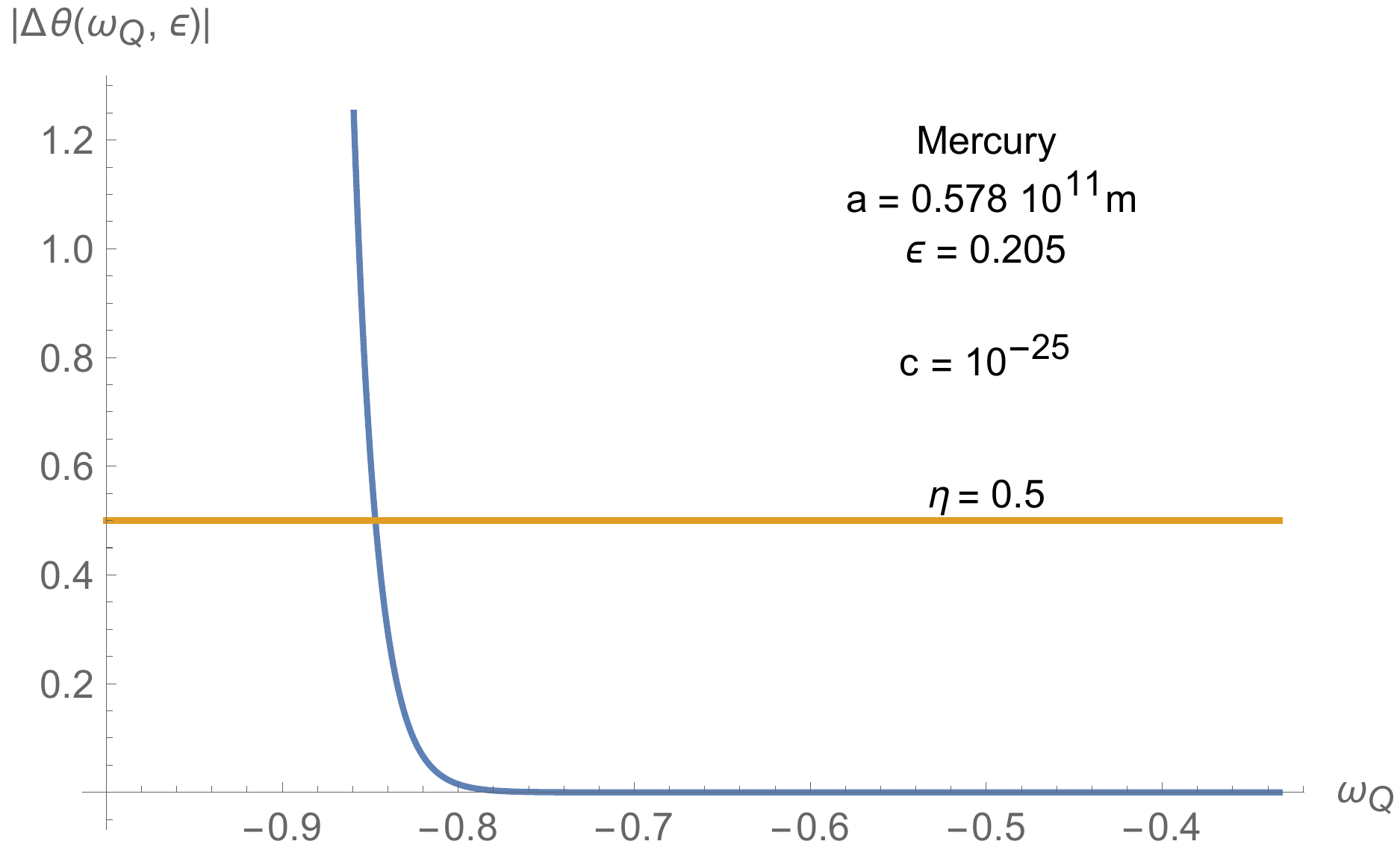}}
    \subfigure[]{\includegraphics[width=0.4\textwidth]{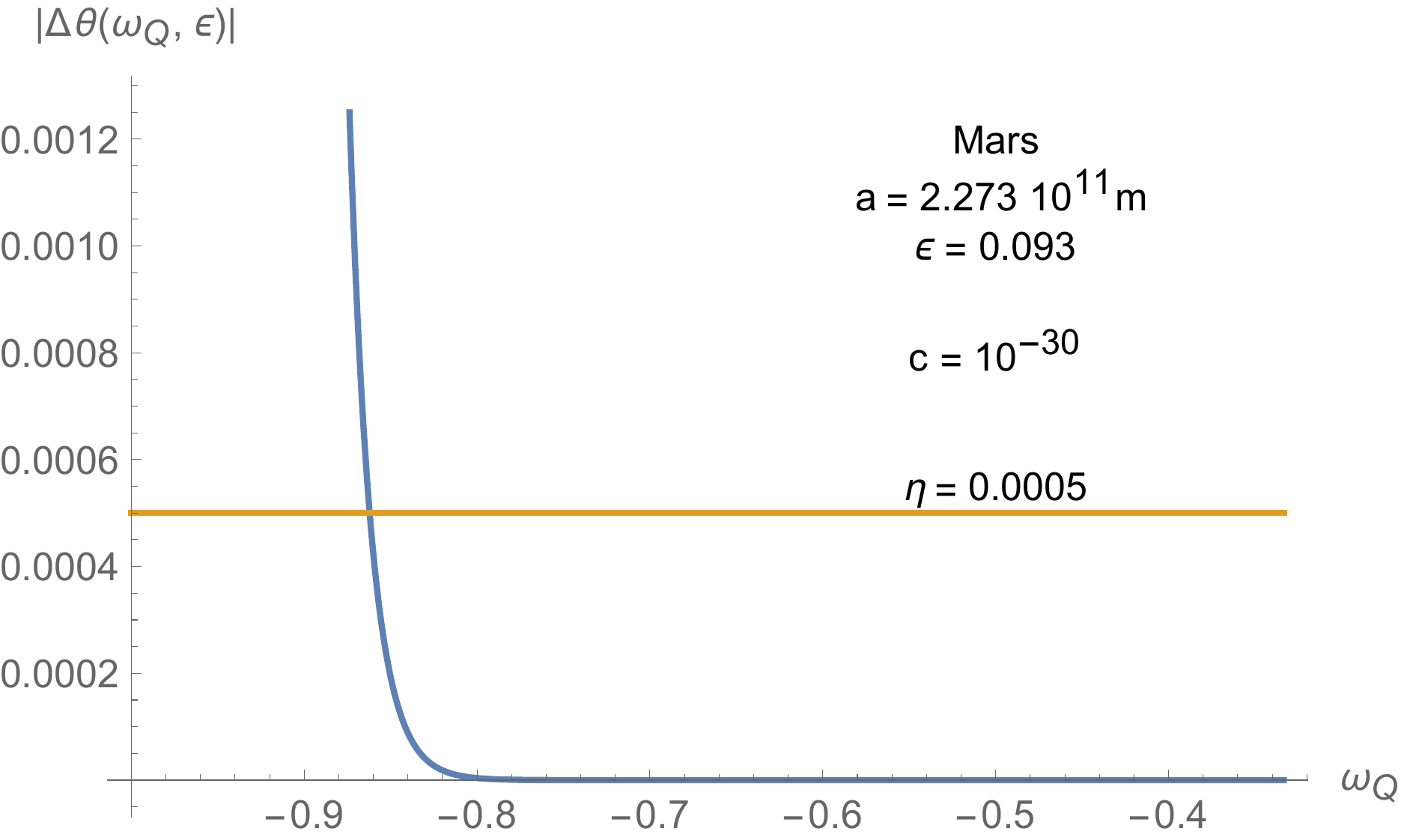}}
    \subfigure[]{\includegraphics[width=0.4\textwidth]{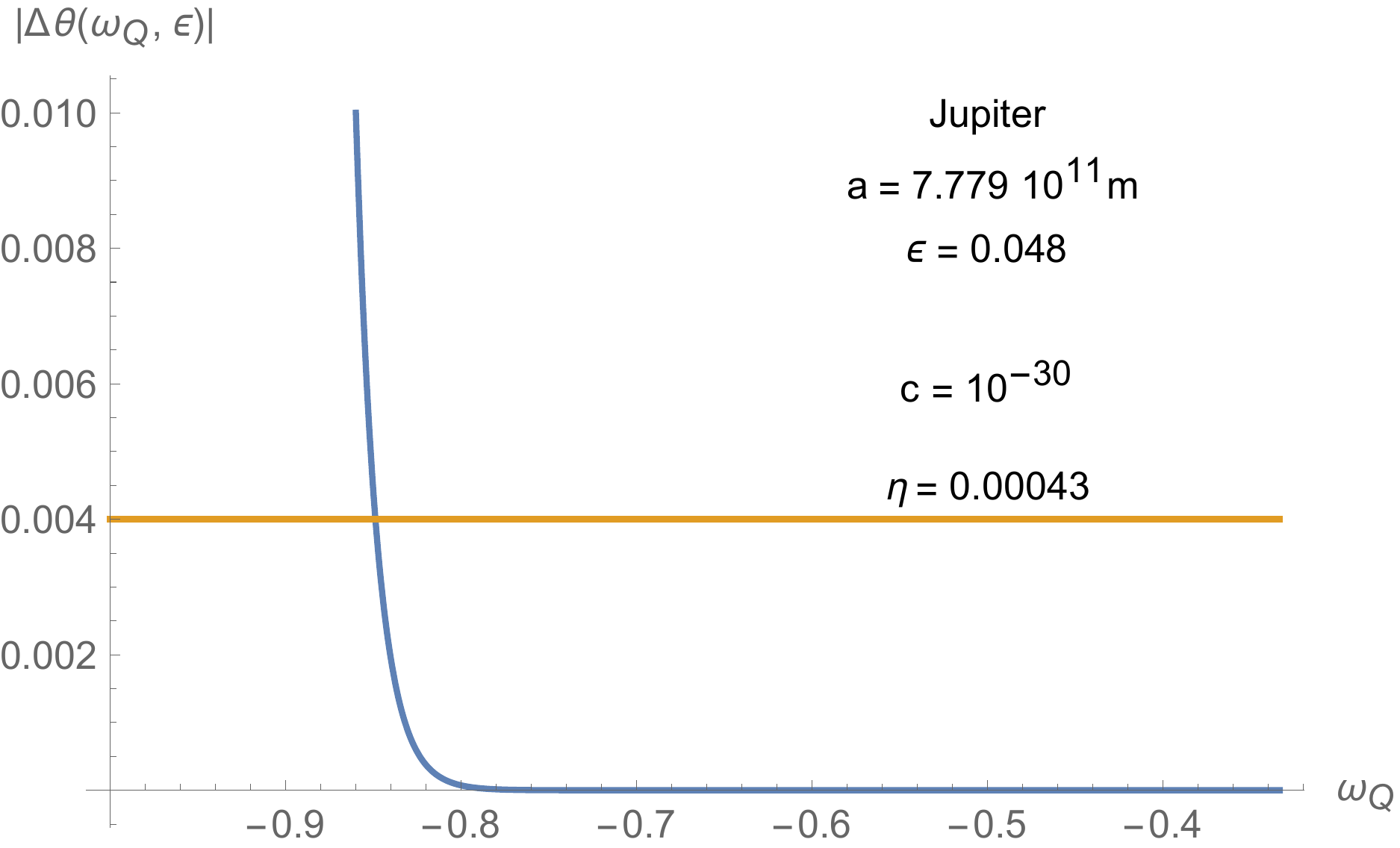}}
    \subfigure[]{\includegraphics[width=0.4\textwidth]{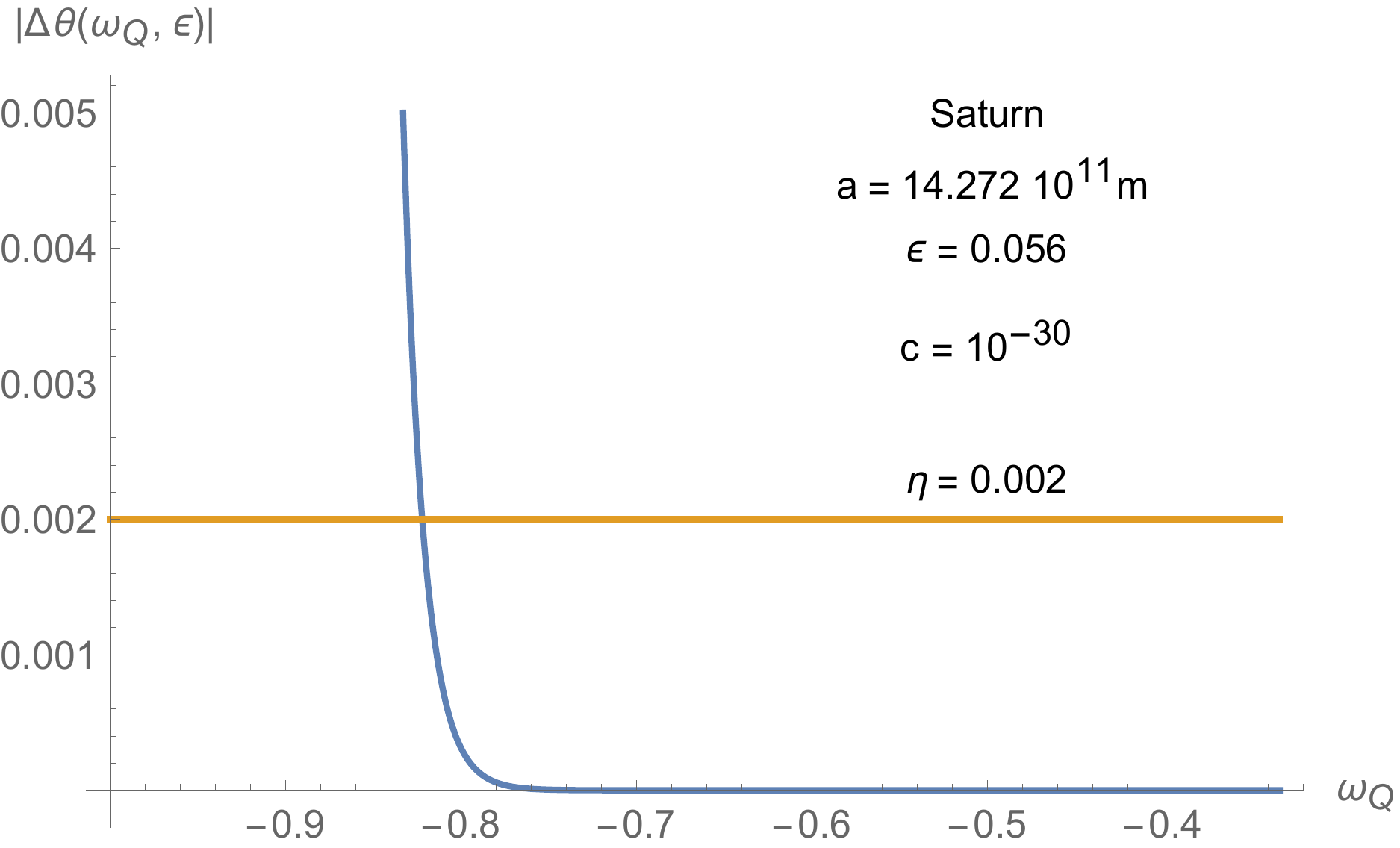}}
    \caption{(a) $|\Delta \theta(\omega_Q, \epsilon)|$ vs $\omega_Q$ for Mercury. (b) $|\Delta \theta(\omega_Q, \epsilon)|$ vs $\omega_Q$ for Mars. (c) $|\Delta \theta(\omega_Q, \epsilon)|$ vs $\omega_Q$ for Jupiter. (d) $|\Delta \theta(\omega_Q, \epsilon)|$ vs $\omega_Q$ for Saturn.}
    \label{figPowLaw}
\end{figure}

An interesting  possibility we wish to discuss is related to quintessence field, invoked to explain the speed-up of the present Universe \cite{Jamil:2014rsa}. Quintessence may generate a negative pressure, and, since it is diffuse everywhere in the Universe, it can be the responsible of the observed accelerated phase, as well as it is present around a massive astrophysical object deforming the spacetime around it \cite{Kiselev:2002dx}.
The studies of quintessential black holes are also motivated from M-theory/superstring inspired models
\cite{Belhaj:2020oun,HeydarFard:2007bb,HeydariFard:2007qs}
(see \cite{Chen:2008ra,Toshmatov:2015npp,Abdujabbarov:2015pqp,Ghosh:2015ovj,Jamil:2014rsa,Belhaj:2020rdb,
Israr AliIJMPD20,Khan5essence,Abbas:2019olp,Javed:2019jag,Uniyal:2014paa,GLNeutrino} for applications). The solution of Einstein's field equations for a static spherically symmetric quintessence surrounding a black hole in 4 dimension is given by \cite{Kiselev:2002dx,Chen:2008ra}
  \begin{equation}
g_{\mu\nu}=\text{diag}\left(-f(r), f^{-1}(r), r^2, r^2 \sin^2 \theta\right)\,,
\label{metric}
\end{equation}
with
\begin{equation}\label{frmetric}
f(r)=1-\frac{2M}{r}-\frac{c}{r^{3\omega_Q+1}}\,,
\end{equation}
where $\omega_Q$ is the adiabtic index (the parameter of equation of state), $-1\leqslant\omega_Q\leqslant -\frac{1}{3}$, and $c$ the quintessence parameter.
The cosmological constant ($\Lambda$CMD model) follows from (\ref{metric}) and (\ref{frmetric}) with $\omega_Q =-1$ and $c=\Lambda/3$,
\begin{equation}
    f(r)=1-\frac{2M}{r}-\frac{\Lambda r^2}{3} \,\ .
\end{equation}
The Quintessential potential reads $V_Q=-\frac{c}{r^{3\omega_Q+1}}$, so that comparing with (\ref{DeltaPowerla}) one gets
 \[
 q\to -(3\omega_Q+1)\qquad \alpha_q \to c\,.
 \]
The precession (\ref{precesspowerlaw}) leads to
\begin{equation} \label{precesspowerlaw2}
|\Delta \theta_p(\omega_Q, \epsilon)| = \frac{\pi c}{G M} a^{-3\omega_Q} \sqrt{1-\epsilon^2} \chi_{\omega_Q}(\epsilon)\,,
\end{equation}
with
\begin{equation}\label{hyper2}
\chi_{\omega_Q}(\epsilon) = 3\omega_Q (1+3\omega_Q) \,\, {_2F_1} \left (\frac{2+3\omega_Q}{2},\frac{3+3\omega_Q}{2}\, ;2\, ;\epsilon^2 \right ) \, .
\end{equation}
By requiring $|\Delta \theta_p(\omega_Q, \epsilon)| \lesssim \eta$ one gets the bounds on the parameters $\{\omega_Q, c\}$. Results are reported in Table \ref{tablePowLaw} and Fig. \ref{figPowLaw} for fixed values of $c$.

\begin{table}
\begin{center}
\caption{Values of the parameter $\omega_Q$ obtained from (\ref{precesspowerlaw2}) using the values of periastron advance for planets of the Solar System.}
\begin{tabular}{|c||c|c|c|}
	\hline
	Planet & $\eta$ &  $c(m^{3\omega_Q+1})\sim$ & $\omega_Q\gtrsim$ \\
	\hline \hline
Mercury & 0.5 & $10^{-25}$ & -0.86 \\
Mars & $5 \times 10^{-4}$ & $ 10^{-30}$ & -0.88 \\
Jupiter & $4\times 10^{-3}$ & $10^{-30}$ & -0.84 \\
Saturn & $2\times 10^{-3}$ & $10^{-30}$ & -0.82 \\
	\hline
\end{tabular}
\label{tablePowLaw}
\end{center}
\end{table}

\section{Test on S2 Star}
\begin{figure}
    \centering
    \subfigure[]{\includegraphics[width=0.4\textwidth]{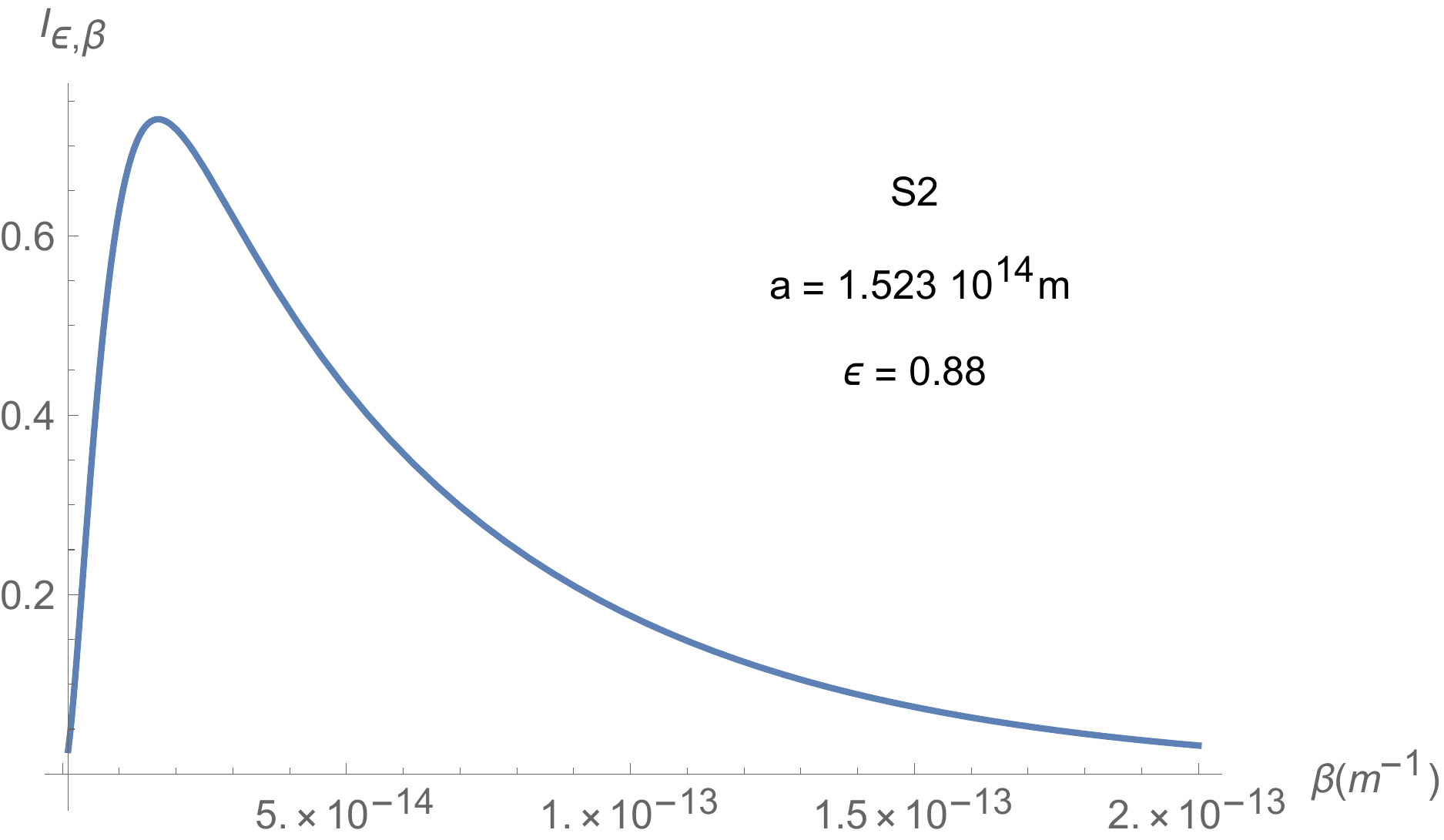}}
    \subfigure[]{\includegraphics[width=0.4\textwidth]{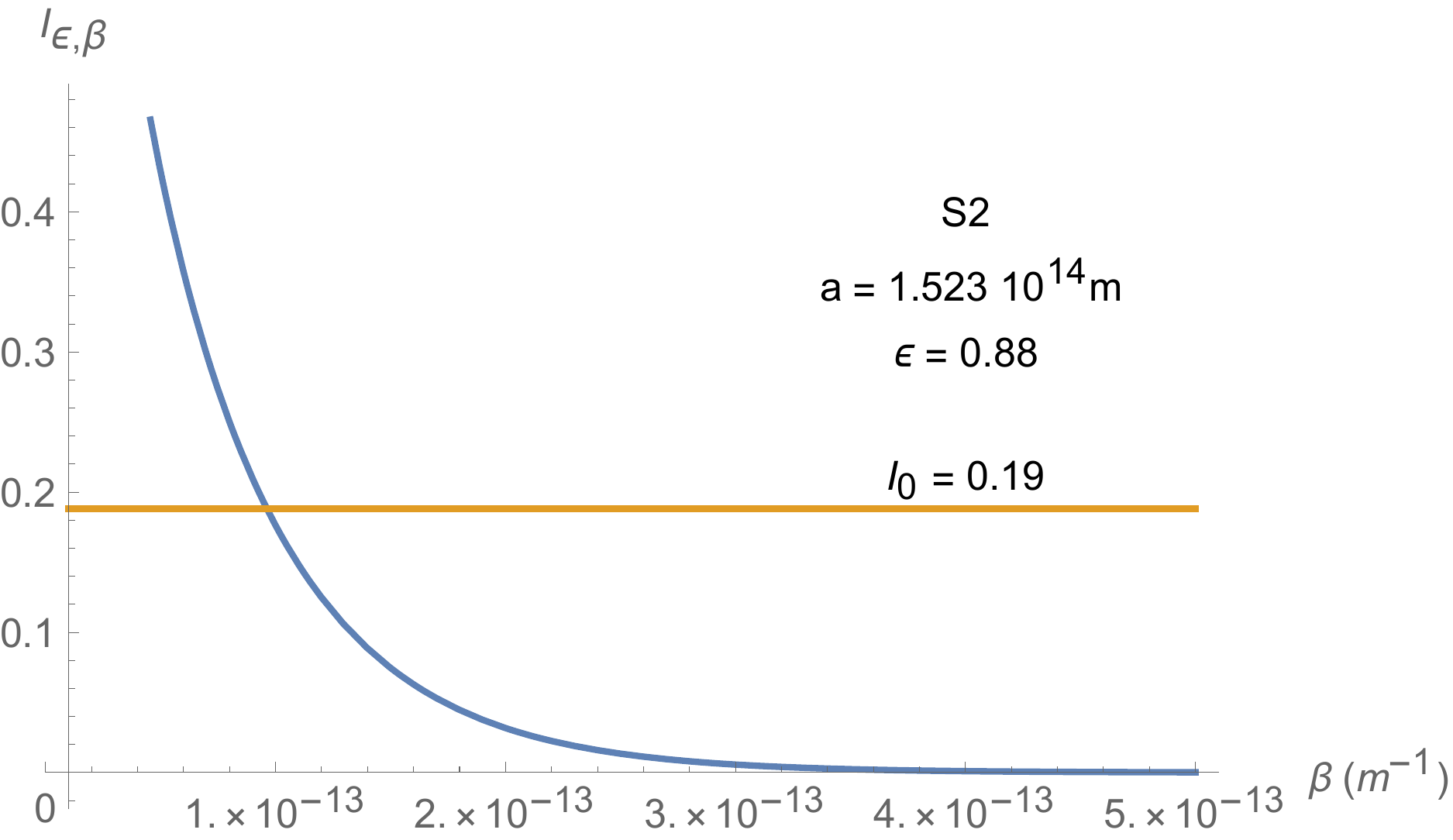}}
    \subfigure[]{\includegraphics[width=0.4\textwidth]{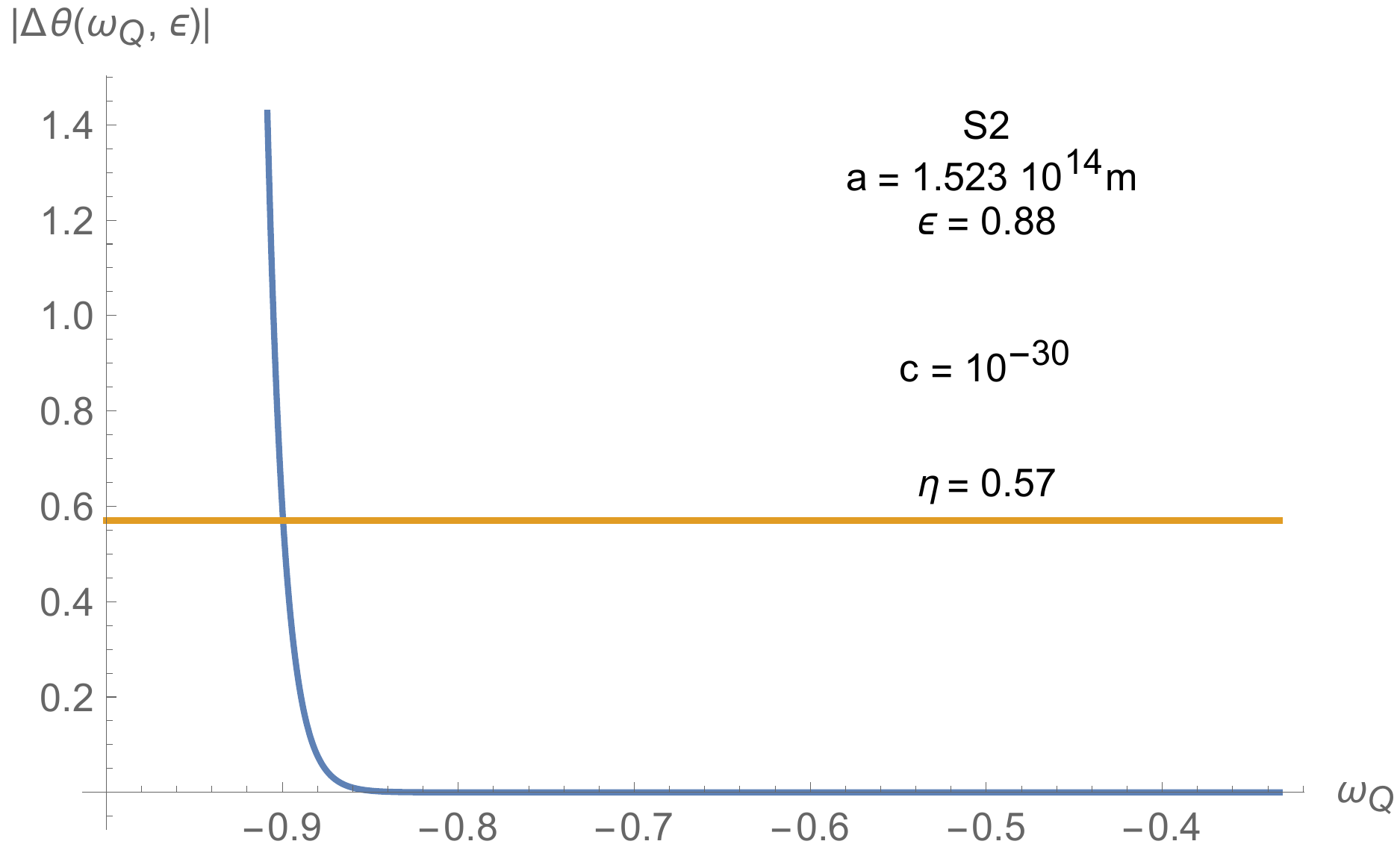}}
    \caption{(a) $I_{\epsilon, \beta}$ vs $\beta$ for S2 star in FOG theories. (b) $I_{\epsilon, \beta}$ vs $\beta$ for S2 star. (c) $|\Delta \theta{\omega_Q, \epsilon}|$ vs $\omega_Q$ for S2 star. ($c$ is in $m^{3\omega_Q+1}$ units).}
    \label{figS2}
\end{figure}

Finally, we shortly conclude our analysis testing the modified gravity predictions for S2 Star orbiting around Sagittarius A*, the Supermassive Black Hole at the center of the Milky Way, which has got a mass equal to $M=(4.5 \pm 0.6)\times 10^{6} M_{\odot}$  and a Schwarzschild radius $R_S = 2GM = 1.27 \times 10^{10} \, m$. The S2 Star orbit has an eccentrity $\epsilon=0.88$ and a semi-major axis $a=1.52917 \times 10^{14}$m. According to Ref. \cite{iorio}, the periastron advance is $(0.2\pm 0.57)$deg, hence $\eta=0.57$ (it is expected that the interferometer GRAVITY may improve such an accuracy level). We discuss the periastron advance for the gravitational models above discussed:

\begin{itemize}
\item {\bf STFOG} - Referring to Scalar-Tensor Fourth Order Gravity, from Eq. (\ref{Fibound}) one gets
  \begin{equation}\label{FiboundS2}
 |\Delta \theta_p(\kappa,\epsilon)|\lesssim \eta \quad \to \quad  |F_i|\lesssim \frac{\eta \epsilon}{2I_{\epsilon, \beta_i}}\sim 0.36 \,, \,\, i=\pm, Y\,.
 \end{equation}
where in Fig. \ref{figS2}(a) is plotted the function $I_{\epsilon, \beta}$ for the S2 star. We ave taken the maximum value of $I_{\epsilon, \beta}$ corresponding to $\beta \sim 2\times 10^{-14} m^{-1}$ (see Fig. \ref{figS2}(a)). The analysis of S2 star orbit around the Galactic Centre in $f(\phi, R)$ and $f(R, \Box R)$ has been investigated in \cite{Borka}.
\item {\bf NCSG} - The S2 star values $\{\epsilon, \eta, a\}$ imply that, from (\ref{betaNCSGbound}),
 \begin{equation}\label{betaNCSGboundS2}
 |\Delta \theta_p(\beta,\epsilon)|\lesssim \eta \quad \to \quad  |I_{\epsilon, \beta}|\lesssim I_0\,, \quad I_0\equiv \frac{3 \eta \epsilon}{8} \simeq 0.19\,.
 \end{equation}
 Results are reported in Fig. \ref{figS2}(b). We can see that the lower bound on $\beta$ is $\beta\gtrsim 1.1 \times 10^{-13}m^{-1}$. These bounds are compatible with the astrophysical bounds \cite{Nelson:2010rt, Nelson:2010ru}.
 \item {\bf Quintessence} - In the case of Quintessence field deforming the Schwarzschild geometry, Eq. (\ref{precesspowerlaw2}) implies
  \begin{eqnarray} \label{precesspowerlawS2}
|\Delta \theta_p(\omega_Q, \epsilon)| &=& \frac{\pi c}{G M} a^{-3\omega_Q} \sqrt{1-\epsilon^2} \chi_{\omega_Q}(\epsilon)\lesssim 0.57 \,, \\
\chi_{\omega_Q}(\epsilon) &=& 3\omega_Q (1+3\omega_Q) \,\, {_2F_1} \left (\frac{2+3\omega_Q}{2},\frac{3+3\omega_Q}{2}\, ;2\, ;\epsilon^2 \right ) \, .
\end{eqnarray}
Results are reported in Fig. \ref{figS2}(c), from which it follows that for Quintessence $|\Delta \theta_p(\omega_Q, \epsilon)|\lesssim 0.57$ provided $\omega_Q \gtrsim 0.9$. Therefore, the exact value $\omega_Q=-1$ corresponding to the cosmological constant is excluded in this range of values.
\end{itemize}

\section{Conclusions}\label{conclusions}

In this paper we have  studied the periastron advance of Solar system Planets in the case in which the gravitational interactions between massive bodies is described by modified theories of gravity. In these models the corrections to the Newtonian gravitational interaction is of the Yukawa-like form, $V(r)=V_N(1+\alpha e^{-\beta r})$, where $V_N=GM/r$ is the Newtonian potential, or the power-law form, $V(r)=V_N+\alpha_q r^q$. To compute the corrections to the periastron advance, we have used results of Ref. \cite{mcdell} in which the general formulas are provided in terms of the central body's mass $M$, and the orbital parameters $a$ and $\epsilon$, the semi-major axis and eccentricity of the orbit, respectively. This two-body system constitutes a good model for many astrophysical scenarios, such as those at the scale of Solar System, constituted by the Sun and a planet, as well as binary system composed by a Super Massive Black Hole and an orbiting star, which are both the most suitable candidates to test a gravitational theory.

In the case of Scalar Tensor Fourth Order Gravity, we find that the parameters of the model are given by (see Eqs. (\ref{betaFOG}, \ref{FparameterFOG}, \ref{alphaFOG}) $\alpha \sim F_i$, $\beta \sim \beta_i$, with $i=\pm, Y$, $F_+ =g(\xi,\eta)\,F(m_+ {\cal R})$, $F_- = \Big[\frac{1}{3}-g(\xi,\eta)\Big]\,F(m_- {\cal R})$, $F_Y=- \frac{4}{3}\,F(m_Y {\cal R})$, $\beta_\pm = m_R \sqrt{\omega_\pm}$, $\beta_Y = m_Y$. The greatest value of $\beta_i$ is
$\beta_i \sim 5\times 10^{-11}m^{-1}$, which leads to the constraint on $F_i$ is $F_i< 10^{-4}$. This allows to get a bound on the massive modes $m_i$, $i=\pm, Y$, corresponding to the extra modes presents in ETG.

In the case of Non-Commutative Spectral Gravity, we have found that the perihelion's shift of planets allows to constrain the parameter $\beta$ at $\beta > (10^{-11} - 10^{-10}) m^{-1}$ (in this theory the parameter $\alpha$ is given and is of the order $\alpha\sim {\cal O}(1)$).
Such a constraint on the parameter $\beta$ improves several order of magnitude ones derived by using pulsar timing $\beta \geq 7.55\times 10^{-13} {m}^{-1}$ \cite{Nelson:2010rt, Nelson:2010ru}. These constraints, however, are weaker compared to the ones obtained from terrestrial experimental data, E\"ot-Wash~\cite{eot} and Irvine~\cite{irvine} experiments is~\cite{kapner}, which give $\beta \gtrsim 10^4 {m}^{-1}$ (a bound on $\beta$ has been derived from Gravity Probe B experiment, giving $\beta> 10^{-6}m^{-1}$ \cite{stabile}).

We have also studied the Quintessence field surrounding a massive gravitational source. In this case the parameter characterizing the gravitational field are the adiabatic index $\omega_Q$ and the quintessence parameter $c$. The analysis shows that $c$ assumes tiny values, as expected, being essentially related to the cosmological constant, while $\omega_Q\gtrsim -(0.9-0.8)$, that is it never assumes the value $-1$ corresponding to the pure cosmological constant.

The case of the S2 Star around Sagittarius A*, the Super Massive Black Hole at the center of the Milky Way, has been also studied. In such a case we have found that for STFOG and NCSG $\beta > 10^{-13} m^{-1}$, a bound compatible with astrophysical constraints, while for the quitessence field we have inferred $\omega_Q \gtrsim -0.9$.

As a final comment, we point out that there could exist screening mechanism effects operating on Earth and Solar System scales \cite{71,72,73,73a,74}, but could not be effective on larger scales, such as the astrophysical scales. Further observations over larger distances could provide limits on both screening mechanisms and higher derivative corrections, in particular on the effective gravitational model here discussed.

\section*{Acknowledgments}
The authors acknowledge the support of  {\it Istituto Nazionale di Fisica Nucleare} (INFN).

\end{document}